\documentclass[preview,5p]{elsarticle}

\usepackage{easyReview}
\newenvironment{hlbreakable}%
{\color{black}}%
{}

\usepackage{url}

\usepackage{stackrel}
\usepackage{bm}
\usepackage{graphics,amssymb,epsfig,color}
\usepackage{xcolor}
\usepackage{graphicx}
\graphicspath{{./images/}{./figures/}{./graphics/}}

\usepackage{colortbl}%

\definecolor{LightCyan}{rgb}{0.5,1,1}
\definecolor{LightMagenta}{rgb}{1,0.6171875,0.9453125}
\definecolor{LightYellow}{rgb}{0.95,0.95,0.5}
\definecolor{LightGreen}{rgb}{0.6,0.95,0.6}
\definecolor{LightRed}{rgb}{0.95,0.5,0.5}

\definecolor{LightGray}{rgb}{0.82421875,0.82421875,0.82421875}
\definecolor{Gray}{rgb}{0.66015625,0.66015625,0.66015625}

\usepackage{booktabs}
\renewcommand{\arraystretch}{1.2}

\usepackage[english]{babel}
\usepackage[autostyle, english = american]{csquotes}
\MakeOuterQuote{"}

\usepackage{makecell}
\usepackage{array}

\newcolumntype{C}[1]{>{\centering\arraybackslash}p{#1}}

\usepackage{float}

\usepackage[hidelinks]{hyperref}
\hypersetup{
	colorlinks=true,
	breaklinks=true,
	urlcolor=Blue,
	linkcolor=Blue,
	pdfpagemode=UseNone,
	bookmarksopen=false, 
	pdftitle={Continuous Generation of Volumetric Images During Stereotactic Body Radiation Therapy Using Periodic kV Imaging and an External Respiratory Surrogate},
	pdfauthor={Matthieu Lafreniere},
	pdfkeywords={Respiratory Motion Modeling, Continuous Volumetric Image Generation, Deformable Image Registration (DIR), Principal Component Analysis (PCA), External Respiratory Surrogate, Radiation Therapy, SBRT, 4DCT, Respiratory Motion, Motion Modeling, Treatment Verification, Dose Assessment, Digitally Reconstructed Radiograph (DRR), Digital Image Reconstruction (DIR), Forward Iterative Reconstruction (FIR), Varian, TrueBeam, Eclipse, XCAT Phantom}
}  
\usepackage[]{cleveref}

\newcommand\figureWidth{1.00}

\begin{document}
	\sloppy

	\title{\begin{hlbreakable}Continuous Generation of Volumetric Images During Stereotactic Body Radiation Therapy Using Periodic kV Imaging and an External Respiratory Surrogate\end{hlbreakable}}
	
	\author[1]{M Lafreni\`ere\corref{cor1}}
	\ead{matthieu_lafreniere@dfci.harvard.edu}
	
	\author[1]{N Mahadeo}
	
	\author[2]{J Lewis}
	
	\author[3]{J Rottmann}
	
	\author[1]{C L Williams}
	\ead{cwilliams@bwh.harvard.edu}

	\cortext[cor1]{Corresponding author. Brigham and Women's Hospital, Dana-Farber Cancer Institute, Harvard Medical School, 75 Francis St, ASBI-L2, Boston, MA, 02215, USA. Tel.: 617-525-7142}
	\address[1]{Brigham and Women's Hospital, Dana-Farber Cancer Institute, Harvard Medical School, 75 Francis St, Boston, MA, 02215, USA}
	\address[2]{University of California, Los Angeles, CA, 90095, USA}
	\address[3]{Paul Scherrer Institute, Forschungsstrasse 111, 5232 Villigen, Switzerland}

	\begin{abstract}
		\begin{hlbreakable}We present a technique for continuous generation of volumetric images during SBRT using periodic kV imaging and an external respiratory surrogate signal to drive a patient-specific PCA motion model. Using the on-board imager, kV radiographs are acquired every 3~seconds and used to fit the parameters of a motion model so that it matches observed changes in internal patient anatomy. A multi-dimensional correlation model is established between the motion model parameters and the external surrogate position and velocity, enabling volumetric image reconstruction between kV imaging time points. Performance of the algorithm was evaluated using 10 realistic e\textbf{X}tended \textbf{CA}rdiac-\textbf{T}orso (XCAT) digital phantoms including 3D anatomical respiratory deformation programmed with 3D tumor positions measured with orthogonal kV imaging of implanted fiducial gold markers. The clinically measured ground truth 3D tumor positions provided a dataset with realistic breathing irregularities, and the combination of periodic on-board kV imaging with recorded external respiratory surrogate signal was used for correlation modeling to account for any changes in internal-external correlation. The three-dimensional tumor positions are reconstructed with an average root mean square error (RMSE) of 1.47 mm, and an average 95$^{\rm th}$ percentile 3D positional error of 2.80 mm compared with the clinically measured ground truth 3D tumor positions. This technique enables continuous 3D anatomical image generation based on periodic kV imaging of internal anatomy without the additional dose of continuous kV imaging. The 3D anatomical images produced using this method can be used for treatment verification and delivered dose computation in the presence of irregular respiratory motion.\end{hlbreakable}
	\end{abstract}


	\maketitle

	\section{Introduction}{\label{sec:Introduction}}	
	Respiratory motion is an outstanding challenge for radiation therapy treatments of many thoracic and abdominal malignancies. This motion may result in organ displacements of several centimeters, particularly in the superior-inferior (SI) direction \cite{AAPMTG76,KorremanIGRTmotionManagement2015,KorremanMotionInRadiotherapy2012}. The respiratory motion is three-dimensional and non-rigid, affecting both the target lesion(s) as well as adjacent organs at risk, potentially causing deviations in the planned dose (target underdosage and increased dose to normal tissues) \cite{JiangMobileTumorsRT2006,AAPMTG76,KorremanIGRTmotionManagement2015,Vedam4DCT2003,GeoffreyTumorMotionEffectOnPlanning2003}.
	
	\begin{hlbreakable}
		In current standard clinical practice, accounting for organ motion relies on the use of 4DCT \cite{Rietzel4DCT2005,Pan4DCT2005} to define an internal target volume that encompasses the target motion. In standard 4DCT acquisition protocols, the imaging represents just one or two respiratory cycles in each region of the patient anatomy. This inherently presents a limited view of the patient's breathing pattern as it does not reflect any breath-to-breath changes in motion. This can lead to substantial uncertainties in the internal target volume derived from a 4DCT dataset \cite{StJames4DCT2012,Sarker4DCT2010,JingCaiGatedRT2010,JingCai4DCT2008}. Moreover, 4DCT images obtained during pre-treatment simulation may not represent internal patient anatomy and organ motion at the time of treatment, as the patient's breathing pattern may change in the intervening time \cite{StJames4DCT2016,Mishra3DGeneration2013}. Even with modern techniques such as using 4DCT during simulation, using breathing coaching or the ABC method, there are still residual uncertainties in the location of the target and normal tissue due to day-to-day changes in a patient's breathing pattern. The impact of these day-to-day changes on the delivered dose is not well known, and may lead to deviations between the planned and the delivered dose. Generating continuous 3D images would allow to assess these changes and to potentially re-calculate the delivered dose, to improve the accuracy and safety of SBRT.
	\end{hlbreakable}
	
	Respiratory motion modeling techniques can be used to overcome some of the limitations of traditional 4DCT-based techniques, enabling improved 3D localization of target lesions and surrounding normal tissues \cite{ShiehBayesianMarkerless3DTracking2017,FassiExternalSurrogate2014,ZhangMotionArtifacts2010,LewisMarkerlessTracking2010,ZhangPatientSpecificMotionModel2007,SohnPCA2005}. Often, these patient-specific motion-models are created with information available prior to radiotherapy treatment, such as 4DCT \cite{HertantoMotionModel2012,LiPCA2011,SohnPCA2005,LowMotionModel2005}, 4DCBCT \cite{CaiDoseAssessment2015,Dhou4DCBCT2015}, or an external surrogate signal \cite{HurwitzExternalSurrogate2015}. Deformable image registration (DIR) can be performed between the 3D images of each respiratory phase to produce a set of 3D displacement vector fields (DVFs) that describe the respiratory motion in these scans.  
	
	A simplified model of the motion-induced deformations can be created by using a dimensionality-reduction technique, such as PCA, which allows the full DVF to be described as a linear combination of a small subset of vector fields. The entire 3D representation of a patient's anatomy at an arbitrary respiratory state can then be generated using only a small number of model parameters. Other information, such as lung tidal volume and airflow \cite{LowMotionModel2005}, or an external respiratory surrogate \cite{HurwitzExternalSurrogate2015,McClellandRespiratorySurrogate2011,Staub4DCBCTPCA2011,ChoXrayTumorTrackingWithSurrogate2008,ZhangPatientSpecificMotionModel2007} can also be incorporated in the model.
	
	3D fluoroscopic image generation is a technique that uses a patient-specific respiratory motion model in combination with 2D x-ray projection images acquired during treatment to reconstruct a full 3D volume \cite{Mishra3DGeneration2013}. Parameters of the motion model are determined using an iterative optimization procedure in which a digitally reconstructed radiograph (DRR) of the modeled patient anatomy is compared with the acquired image, and the model parameters are then adjusted for an optimal match.
	
	\begin{hlbreakable}Previous motion modeling-based approaches have demonstrated that kV or MV images \cite{HarrisKVimagesTo4DMRI2018,CaiDoseAssessment2015,Dhou4DCBCT2015,MishraMVimages2014,Mishra3DGeneration2013}, as well as external surrogate signals \cite{HurwitzExternalSurrogate2015,LiSurrogateBasedVMAT2012}, can be used independently to generate volumetric images. However, these techniques have limitations that can pose challenges for clinical implementation. For kV image-based techniques, the image acquisition frame rate is impacted by the potential for excess imaging dose (continuous fluoroscopy can add up to 1.2 Gy over a treatment course \cite{NgContinuousKVimagingDose2013,CrockerContinuousKVimagingDose2012,KeallIGART2015}), as well as machine constraints (e.g. the Varian TrueBeam platform used in our clinic limits in-treatment kV imaging to a rate of 1/3 Hz in clinical mode), which may result in images being acquired too infrequently to fully capture the motion during each breath. The use of MV portal images is also limited due to the small and complex apertures used with modulated treatments as well as reduced soft-tissue contrast. For external surrogate-based techniques, the correlation between the surrogate signal and internal patient anatomy can change over time \cite{WuExternalSurrogate2008,KanoulasExternalSurrogate2007,BerbecoExternalSurrogate2005}.
		
		Prior studies have shown that it is possible to track tumor motion using orthogonal kV imaging of implanted fiducial gold markers and with an external surrogate infrared signal to assess internal-external anatomy correlation \cite{AdlerOrthogonalXRayAndExternalSurfaceMonitoringCyberknife2000,YangXLTS4mm2017,TorshabiFuzzyModel2013,CollinsCyberknifeRealTimeTumorMotionTracking2007,SayehXLTS1p5mm2007,FuXLTS1p5mm2007,KilbyCyberKnife2010}.  However, these techniques only track the motion of a the fiducial markers, and do not capture the full three-dimensional motion, which may include relative motion between the tumor and nearby organs at risk.  Furthermore, many clinical accelerators often do not have fixed orthogonal x-ray systems.
	\end{hlbreakable}
	
	We present a technique that combines periodic triggered kV images with an external respiratory surrogate to drive a patient-specific motion model built from a pre-treatment 4DCT scan. This method is novel in that it combines image-derived model fitting with a continuous external surrogate signal, providing the capability to generate continuous volumetric images based on kV images of internal anatomy changes while at the same time requiring only limited kV imaging.  This technique addresses many limitations of previous motion modeling techniques and enables high time-resolution 3D anatomical image generation using information that is available on current clinical treatment machines and often already acquired as part of current clinical treatments.

	\section{Methods and materials}{\label{sec:Methods}}
	
	\subsection{Technique Overview}
	This technique extends prior work that generates 3D anatomical images based on single kV radiographs \cite{Mishra3DGeneration2013,MishraMVimages2014} or external surrogate signals \cite{HurwitzExternalSurrogate2015}, and establishes a framework that uses information from both approaches through a correlation model. \begin{hlbreakable}It is designed to rely on capabilities that are already present in or integrated with many current clinical treatment machines: an on-board kV imaging source and panel, and an external respiratory surrogate.
		
		A patient would be set up for treatment according to clinical standard of care, with the addition of a respiratory monitoring device such as an abdominal marker block (e.g. as in the RPM system \cite{AAPMTG76,KuboBSRT2000,FordRPM2002522,WagmanRPM2003659,KeallRPM200481,GiraudRespiratoryGatedRadiotherapy2005,GiraudRespiratoryGating2010}) or spirometry device \cite{LowMotionModel2005,LowSpirometry4DCT2003,ZhangSpirometerBasedPlanOptimization2004,ZhangSpirometerBasedDIBH2003,ZhangSpirometryPatientCoaching2003}. Additionally, during treatment the kV imaging panel would be deployed and would acquire kV images every 3 seconds (this is the current maximum frame rate of the in-treatment kV image functionality in clinical mode on the Varian TrueBeam accelerators used in our clinic).  The information from these modalities will then be combined with a motion model from a treatment planning 4DCT (or potentially a pre-treatment 4DCBCT) to enable volumetric image generation.
	\end{hlbreakable}

	\subsection{Modified digital XCAT phantom with realistic breathing from clinically measured patient data}
	\label{XCAT}
	
\begin{figure}[!htb]
		\centering
		\includegraphics[width=\figureWidth\linewidth]
		{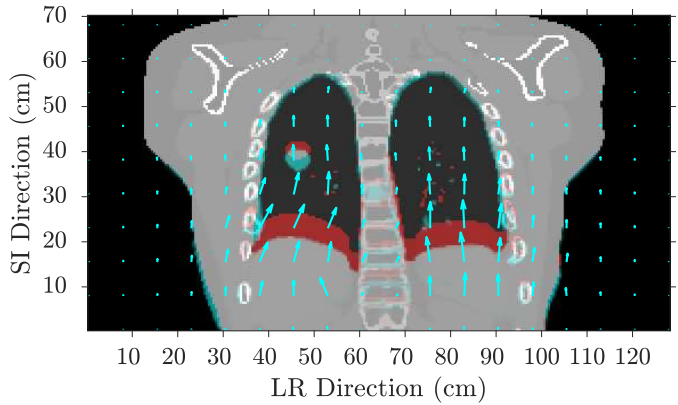}
		\caption[]{Example digital XCAT phantom reference coronal slice image (red) overlayed with a reconstructed coronal slice image from an inhale respiratory phase (cyan) of phantom \#5, showing the tumor and diaphragm moving inferiorly. The reconstructed displacement vector field between the reference and reconstructed image is also shown (cyan arrows).}
		\label{DVF}
	\end{figure}

	The overall accuracy of a time-varying volumetric anatomical image generation technique can be difficult to assess with clinically acquired data due to the lack of a ground truth volumetric image to compare the generated volumetric image against. To address this issue, we use the hybrid digital e\textbf{X}tended \textbf{CA}rdiac-\textbf{T}orso (XCAT) phantom \cite{SegarsXCAT2010,SegarsThesis,SegarsMCAT2001} seen in \cref{DVF}. This phantom incorporates realistic anatomy \cite{SpitzerVisibleHumanDataset1998} for which a ground truth is unambiguously defined from clinical tumor position measurements, and also allows for customized irregular breathing profiles \cite{BerbecoFluoroscopicGating2005,MishraModifiedXCAT2012}. This technique has been developed and used previously in our group \cite{Mishra3DGeneration2013,HurwitzExternalSurrogate2015,MishraMVimages2014,Dhou4DCBCT2015,CaiDoseAssessment2015,CaiDoseAssessmentSBRT2015,Williams4DXCAT2013}, and has proved useful in quantitatively assessing the accuracy of the 4D anatomical image generation over a variety of patient-simulated conditions.
	
	The tumor motion trajectories that were used to generate the XCAT phantoms were acquired with the Mitsubishi Real Time Radiation Therapy (RTRT) system at the Radiation Oncology Clinic at the Nippon Telegraph and Telephone Company Hospital in Sapporo, Hokkaido, Japan \cite{WuExternalSurrogate2008,KanoulasExternalSurrogate2007,BerbecoExternalSurrogate2005,ShiratoTreatmentPlanning2000,ShiratoHokkaidoDataset2000}. A fluoroscopic x-ray system was used to obtain stereoscopic images to localize 1.5 mm gold fiducial markers implanted in lung tumors of patients receiving radiotherapy. Volumetric digital XCAT phantoms were generated for each measured 3D tumor motion trace. The simulated tumors were modeled as 2 cm diameter spheres whose centroid moves in accordance with the tumor motion measured using the RTRT system. The breathing motion of the XCAT phantom is controlled by the diaphragm displacement and anterior motion of the ribs, and these parameters were varied to reproduce the observed 3D tumor motion.  The remainder of the anatomy is moved and displaced according to the deformation model within the XCAT software, which can be nonlinear.  
	
	In addition to the stereoscopically-acquired tumor positions, the RTRT system also included an external respiratory surrogate (\begin{hlbreakable}the Anzai 733V external respiratory surrogate laser-based system \cite{BerbecoAnzai733V2010,BerbecoExternalSurrogate2005}\end{hlbreakable}) that reported the position of the patient's abdomen at a rate of 30 Hz \cite{ShiratoTreatmentPlanning2000,ShiratoHokkaidoDataset2000}. The synthetic XCAT images produced in this study are thus independent from the external surrogate information used to build the correlation model and being drawn from the RTRT dataset. 
	
	\begin{figure}[!htb]   
		\centering 
		\includegraphics[width=\figureWidth\linewidth]
		{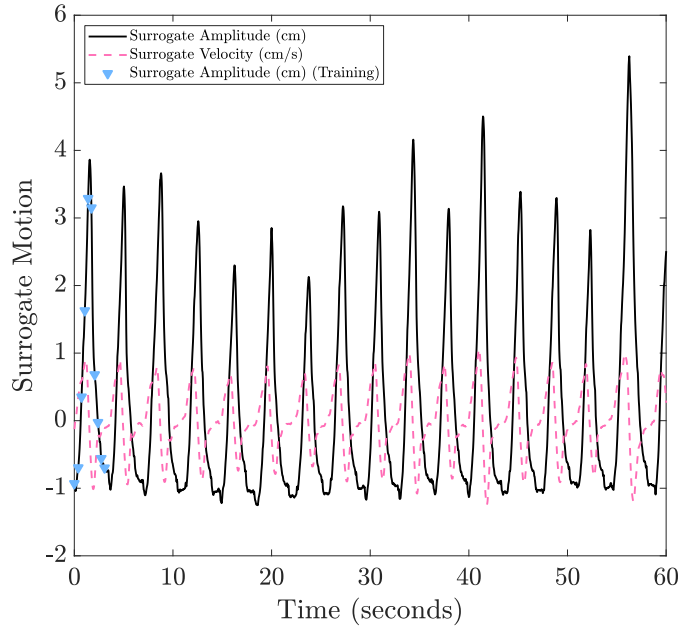}
		\caption[]{First 60 seconds of the respiratory trace from phantom \#5 showing the external surrogate amplitude (plain black line), the surrogate velocity (dashed magenta line) and the time points chosen for generating the simulated 4DCT training phantoms (cyan triangles) used for creating the motion model.}   
		\label{surrogateTrace}
	\end{figure}
	
	Ten XCAT phantom sets were generated from 9 separate patients (two separate datasets came from the same patient). An open source graphical user interface software was used to facilitate the creation of digital XCAT phantoms based on tumor positions clinically recorded with the RTRT system \cite{MyronakisXCATinterface2017}. Phantoms were created with an intrinsic resolution of 2 mm in the left-right (LR) and anterior-posterior (AP) directions, and 2.5 mm in the superior-inferior (SI) direction. The total number of voxels per image is $256 \times 256 \times 140 = 9,175,040$ (LR, AP, SI). The XCAT images were produced every 0.2 second, and for each patient the entire available respiratory trace was used, which varied between 46.7 seconds and 292.0 seconds. 
	
	A training 4DCT dataset consisting of 10 images approximately distributed over the first breathing cycle in the respiratory trace was produced for each XCAT dataset, replicating the information that would be available to train a motion model from clinical 4DCT images (empty cyan triangles in \cref{surrogateTrace}). These "4DCT" XCAT images are spaced every 0.4 seconds for a total of 4 seconds, similarly to previous studies on 4DCT motion modeling using the XCAT phantom \cite{Li3DtumorLoc2011,Mishra3DGeneration2013,MishraMVimages2014}.

	\subsection{Volumetric image generation}
	
	\subsubsection{PCA based respiratory motion model.} 
	Each respiratory phase of the digital XCAT phantom training 4DCT dataset was deformably registered to a peak-exhale reference image from the same dataset using a graphics processing unit (GPU) accelerated double force demons algorithm \cite{GuDemonsDIR2010} with the $\alpha$ parameter fixed to 2.0. A patient-specific motion-model was then constructed by performing PCA on the DVFs obtained with the registration, and retaining only the most important eigenvectors. DVFs describing the respiratory motion can typically be well represented by a linear combination of 2 to 3 PCA eigenvectors \cite{HurwitzExternalSurrogate2015,MishraMVimages2014,Mishra3DGeneration2013,ZhangPCA2013,LiPCA2011,Li3DtumorLoc2011,Staub4DCBCTPCA2011,LiVolumetricReconstruction2010,Vaman4DCTregistration2010,ZhangMotionArtifacts2010,ZhangPatientSpecificMotionModel2007,ZengTomographicRegistration2007}:
	
	\begin{equation}
	\mathbf{DVF} = \overline{\mathbf{DVF}} + \sum_{n=1}^{N} w_n(t) \mathbf{u_n} 
	\label{DVFequation}
	\end{equation}
	
	\noindent where $\overline{\mathbf{DVF}}$ is the mean $\mathbf{DVF}$, $\mathbf{u_n}$ is the $n^{th}$ basis eigenvector, $w_n(t)$ is the time dependent weighting coefficient (also called PCA eigenvalue, or PCA weight) of the $n^{th}$ PCA mode, and $N$ is the total number of PCA modes that are used for motion modeling. Three PCA weighting coefficients were used for each XCAT dataset to build the respiratory motion models for this work.

	\subsubsection{Optimization-based 3D image generation.}
	\label{optimizationImageEstimation}
\begin{figure}[!htb]   
		\centering 
		\includegraphics[width=\figureWidth\linewidth]
		{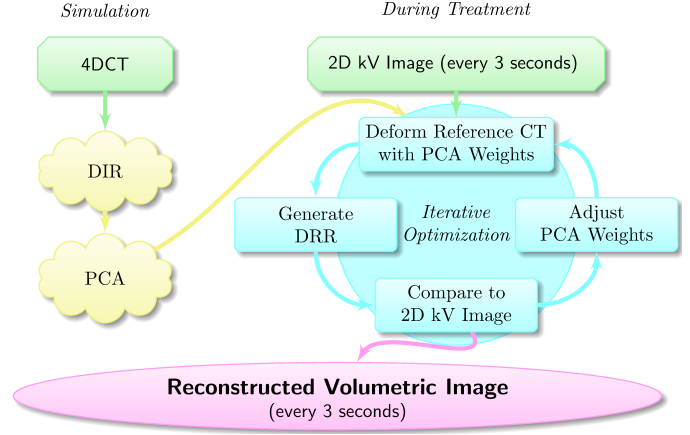}
		\caption[]{Forward iterative reconstruction (FIR) procedure for estimating time-varying volumetric images from single x-ray projections. In this work, a method was developed to use an external respiratory surrogate to generate volumetric images in the 3-second time interval elapsed between each acquired kV image.}   
		\label{4D_image_estimation}
	\end{figure}

	A 3D image can be produced from a single 2D x-ray image by using a forward iterative reconstruction (FIR) process to optimize the weights ($w_n(t)$) in the PCA respiratory motion model so that the projection of the deformed reference CT image matches the observed x-ray image. This deformed reference CT image ($\mathbf{f}$) is created by deforming the reference image ($\mathbf{f_0}$), with a displacement vector field ($\mathbf{DVF}(w)$) determined by the PCA weights. A simulated x-ray image ($\mathbf{P \cdot f}$) can be produced by applying a projection matrix ($\mathbf{P}$) to the deformed reference CT image ($\mathbf{f}$). An iterative  minimization procedure is then performed to match the simulated x-ray image ($\mathbf{P \cdot f}$) and the observed x-ray image ($\mathbf{x}$) acquired during treatment delivery:
	\begin{equation}
	\stackrel[w]{}{min} J(w) = 
	\stackrel[w]{}{min} \| \mathbf{P} \cdot \mathbf{f}\left( \mathbf{DVF}(w), \mathbf{f_0} \right) - \mathbf{\bm\lambda} \cdot \mathbf{x} \| ^2_2,
	\label{gradientDescent}
	\end{equation}
	
	\noindent where $\mathbf{\boldsymbol\lambda}$ is a parameter defining the relative pixel intensity between simulated x-ray projections and acquired x-ray images, and $J(w)$ is a cost function representing the L2-norm (squared least squares) error between the simulated and acquired x-ray image. A version of gradient descent is used to minimize the cost function $J(w)$ (\cref{gradientDescent}). A detailed description of the PCA coefficients optimization can be found in the work of Li \textit{et al.} \cite{Li3DtumorLoc2011,LiVolumetricReconstruction2010}. The resulting deformed reference CT image ($\mathbf{f}$) corresponds to the optimized estimate of the 3D anatomy at the time of the x-ray acquisition. A schematic representation of this process is shown in \cref{4D_image_estimation}. This process is implemented in a GPU-accelerated framework so that it can be performed in real time.
	
	\subsubsection{Continuous volumetric image generation with an external surrogate respiratory trace.}
	During radiotherapy treatments, it may not be possible or desirable to acquire kV images continuously due to imaging dose or machine limitations. Thus, the framework presented in \cref{4D_image_estimation} is extended to continuously generate 3D images in the time interval between each kV image acquisition. To accomplish this, we use the information provided by an external respiratory surrogate to estimate the motion model PCA weights during the image acquisition 3 seconds time interval. This is a non-invasive, non-ionizing modality that provides the amplitude of a patient's abdominal motion with high time resolution (30 Hz).
	
\begin{figure}[!htb]   
		\centering 
		\includegraphics[width=\figureWidth\linewidth]{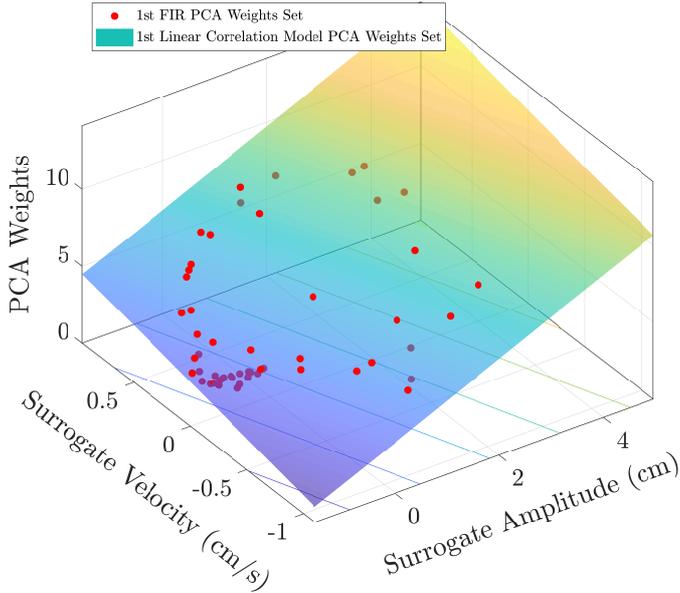}
		\caption[]{The weights of the first principal component of phantom dataset \#5 plotted as a function of the external surrogate amplitude and velocity. The linear correlation model is represented by the fitting surface.}
		\label{correlationModel}
	\end{figure}

\begin{figure}[!htb]
		\centering
		\includegraphics[width=\figureWidth\linewidth]
		{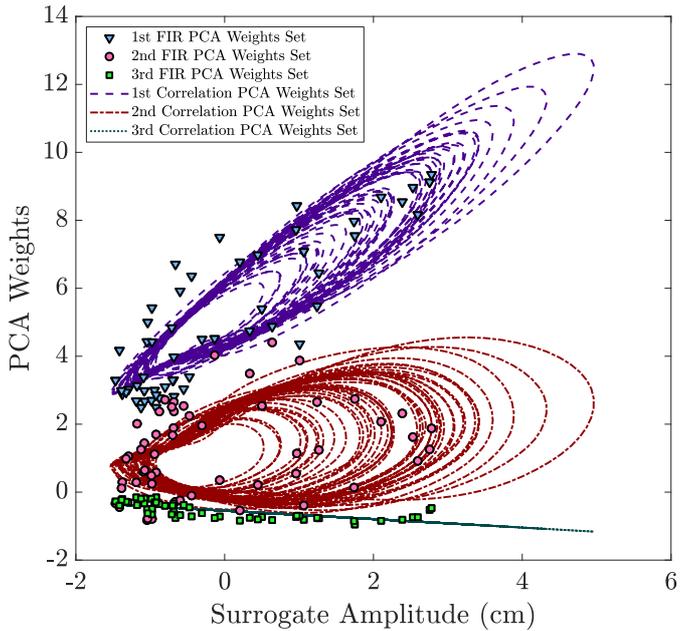}
		\caption[]{Three PCA weights sets created from kV images (markers) and generated with the correlation model (lines) as a function of the external surrogate amplitude for phantom dataset \#5.}    
		\label{PCAvsSurrogateAmplitude}
	\end{figure}

\begin{figure}[!htb] 
		\centering 
		\includegraphics[width=\figureWidth\linewidth]
		{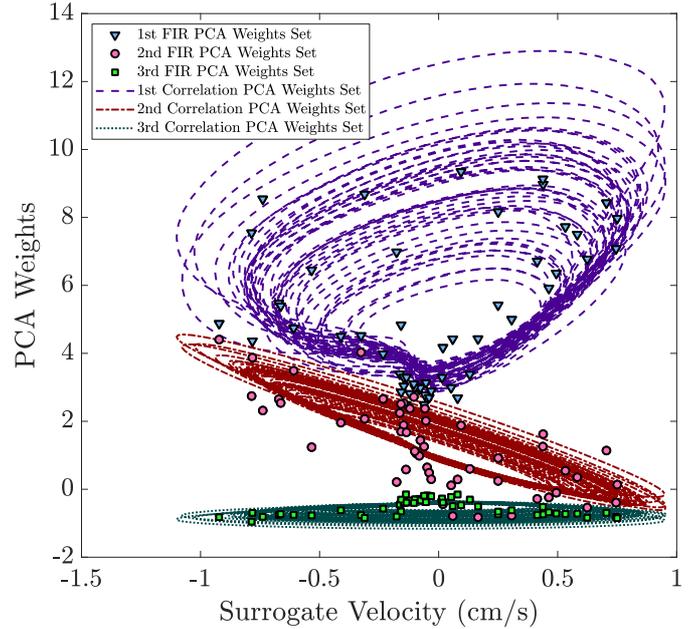}
		\caption[]{{Three PCA weights sets created from kV images (markers) and generated with the correlation model (lines) as a function of the external surrogate velocity for phantom dataset \#5.}}    
		\label{PCAvsSurrogateVelocity}
	\end{figure}

\begin{figure}[!htb] 
		\centering 
		\includegraphics[width=\figureWidth\linewidth]
		{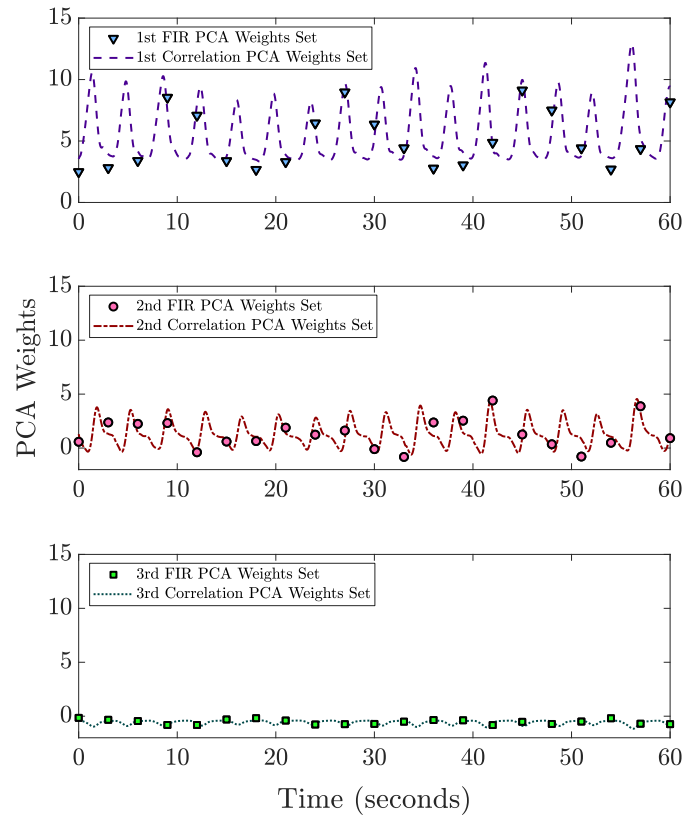}
		\caption[]{{Three PCA weights sets created from kV images (markers) and generated with the correlation model (lines) as a function of time (phantom dataset \#5). The simulated kV image rate used in the motion model for reconstruction is 1/3 Hz. The first PCA component has the dominant motion contribution.}}   
		\label{PCAvsSurrogateVelocityIndividual}
	\end{figure}

	A linear correlation model is developed between the PCA weights, and the displacement and velocity of the surrogate signal. These surrogate characteristics have been shown to be complementary in characterizing respiratory motion \cite{Zhao5DlungMotionModel2009,HurwitzExternalSurrogate2015}. For each kV image acquisition, the weight of each PCA mode, $w_n$, is determined by the procedure described in \cref{optimizationImageEstimation}, and the correlation model is established:
	
	\begin{equation}
	w_n(t) = c_n + f_na(t) + g_n\dot{a}(t)
	\label{correlationEquation}
	\end{equation}
	
	\noindent where $a(t)$ is the amplitude of the external surrogate signal, and $\dot{a}(t)$ is its velocity. The parameters $c_n$, $f_n$ and $g_n$ are specific to each PCA mode of each patient's motion. The linear relationship between the first principal component ($w_1$) and the external surrogate amplitude and velocity ($a(t)$ and $\dot{a}(t)$) is shown in \cref{correlationModel}, and is represented by the linear fitting surface. The first PCA component has the most importance and primarily drives the motion model, as seen in \cref{PCAvsSurrogateAmplitude,PCAvsSurrogateVelocity,PCAvsSurrogateVelocityIndividual}. Correlation model generation is performed on a per-fraction basis because of potential inter-fraction changes in the correlation between internal and external anatomy \cite{WuExternalSurrogate2008,KanoulasExternalSurrogate2007,BerbecoExternalSurrogate2005}.
	
	PCA weights periodically reconstructed with the FIR process and PCA weights continuously generated with the correlation model are shown as a function of the external surrogate amplitude for phantom dataset \#5 in \cref{PCAvsSurrogateAmplitude}. The same PCA weights, but as a function of the external surrogate velocity, are shown in \cref{PCAvsSurrogateVelocity}. Finally, the PCA weights as a function of time are shown in \cref{PCAvsSurrogateVelocityIndividual}.
	
	\subsection{Evaluation criteria}
	\label{evaluationCriteria}
	The tumor localization error is evaluated with the 95$^{\rm th}$ percentile and the RMSE between the tumor centroid location in the reconstructed volumetric images and the ground truth volumetric images. The tumor localization error is estimated in the LR, AP, and SI directions, as well as an absolute 3D positional difference, which provides a geometric measure of the reconstruction accuracy.
	
	As a benchmark for comparison, the FIR process was also performed every 0.2 seconds, to simulate continuous kV fluoroscopy for generating motion model reconstructions.  Comparing these results to the correlation model (which uses imaging every 3 seconds) allows the accuracy of the correlation method to be assessed in the context of the overall performance of the image-based motion modeling techniques.
	
	\section{Results}{\label{sec:Results}}
	The correlation method's performance was evaluated by comparing the position of the tumor centroid in reconstructed images to the tumor centroid position in the ground truth phantom images for each dataset. Simulated kV images were computed every 3 seconds to match the acquisition rate of current radiotherapy clinical systems. The image-based fitting procedure described in \cref{optimizationImageEstimation} was used to generate periodic PCA weights, and continuous correlation model PCA weights were determined using the full external surrogate respiratory trace.  
	
\begin{figure}[!htb]
		\centering 
		\includegraphics[width=\figureWidth\linewidth]
		{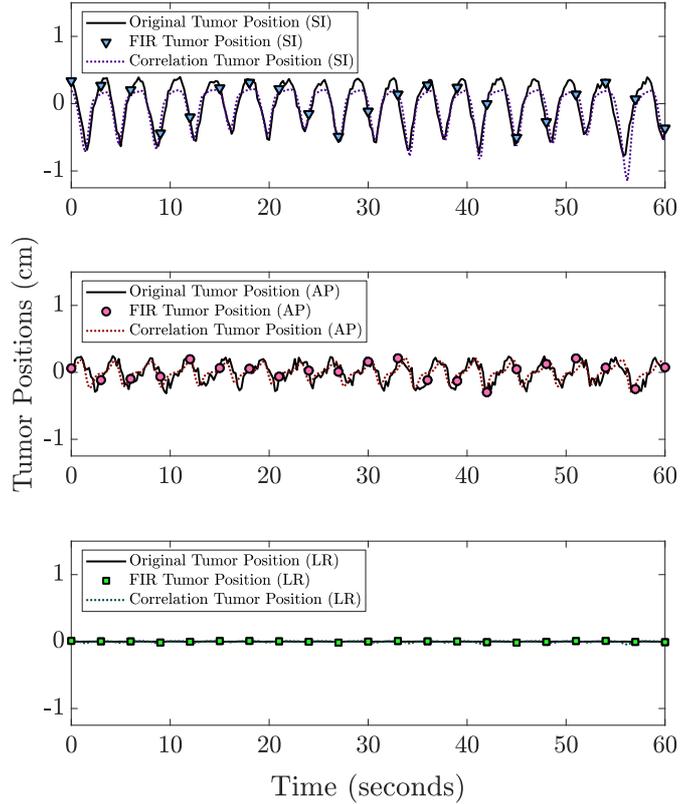}
		\caption[]{{Ground truth tumor positions in the LR, AP and SI directions (plain black lines), alongside those created by the FIR algorithm (markers) and by the correlation model (non-continuous colored lines) (phantom dataset \#5). The simulated kV image rate used in the motion model for reconstruction is 1/3 Hz. The main contribution to respiratory motion comes from the SI direction.}}   
		\label{TumorPositionsIndividual}
	\end{figure}

	Using the correlation model, three-dimensional reconstructions were computed every 0.2 seconds over the entire respiratory trace duration. The average trace length was 155.8 seconds long (ranging from 46.7 to 292.0 seconds). The motion in the datasets was primarily in the SI direction, and the average 3D motion amplitude was 1.7~cm. The ground truth 3D tumor positions compared with the reconstructed 3D tumor positions for a representative respiratory trace are shown in \cref{TumorPositionsIndividual}.
	
\begin{figure}[!htb]
		\centering 
		\includegraphics[width=\figureWidth\linewidth]
		{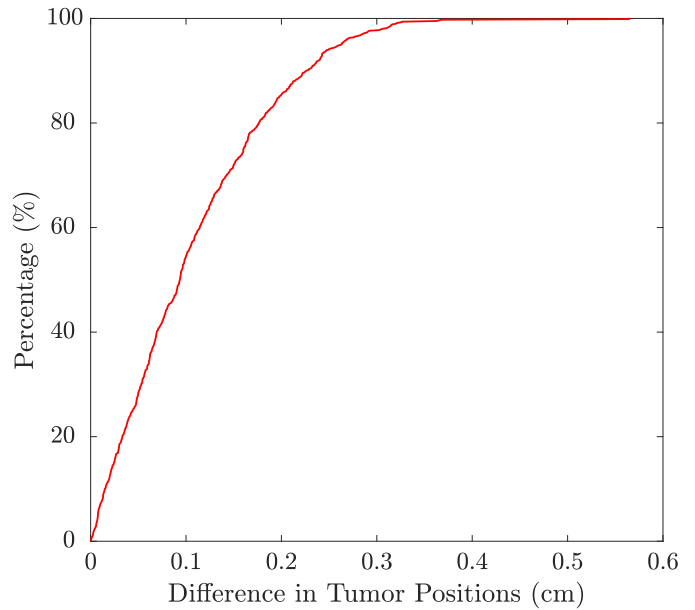}
		\caption[]{{Cumulative difference between the ground truth 3D tumor positions and the 3D tumor positions reconstructed using the correlation model (phantom dataset \#5). For this dataset, the 95$^{\mathbf{th}}$ percentile is 2.62~mm, and the RMSE is 1.36~mm. When averaged over all datasets, the 3D tumor position difference is below 2.80~mm for 95\% of the data, and below 3~mm for 94.6\% of the data.}}    
		\label{reconstructionDifferenceTumorPositions3D}
	\end{figure}

	The accuracy of the 3D tumor positions and the 3D images reconstructed with FIR and the correlation model (COR) for each phantom dataset is summarized in \cref{resultsRMSE}. The average 95$^{\rm th}$ percentile 3D error between correlation model-generated volumetric images and ground truth volumetric images was 2.80~mm, and the RMSE was 1.47~mm (results averaged over 10 different phantom datasets). On average, tumor positions reconstructed using the correlation model method were within 3~mm of the ground truth position 94.6\% of the time, and below 5~mm 99.1\% of the time. A plot of the cumulative 3D differences in tumor location between the ground truth and the correlation model for all generated images from dataset \#5 is shown in \cref{reconstructionDifferenceTumorPositions3D}. Comparison of the full volumetric ground truth and correlation model generated images gives a NRMSE of 97.83\% (refer to \cref{resultsRMSE}). 
	
	A baseline accuracy for the image-based reconstruction method was assessed by performing the FIR process every 0.2 seconds to simulate near-continuous kV acquisition.  This kV-based motion modeling technique was characterized by an average 95$^{\mathbf{th}}$ percentile accuracy of 1.58 mm and an average RMSE of 0.94 mm for the three dimensional tumor position reconstruction. This shows that the tumor position reconstruction obtained with the FIR process is better than a millimeter. When comparing the ground truth and FIR generated volumetric images, the NRMSE is 98.2\% (refer to \cref{resultsRMSE}). 
	
	\begin{figure}[!htb]
		\centering 
		\includegraphics[width=\figureWidth\linewidth]
		{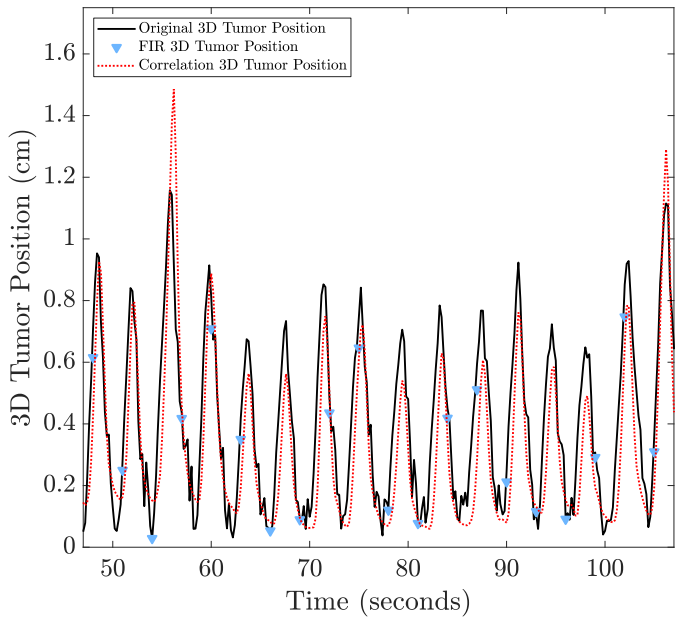}
		\caption[]{{The ground truth three dimensional tumor positions (plain black curve) is displayed alongside the correlation model reconstructed 3D tumor positions (dashed red curve), and the positions determined from performing the FIR process on 1/3~Hz kV images (cyan triangles) (phantom dataset \#5), illustrating the ability of the technique to reconstruct motion with high time resolution, even in the presence of irregular breathing.}}   
		\label{TumorPositions3D}
	\end{figure}
	
	\renewcommand{\arraystretch}{1.35}
	\begin{table*}[!ht]
		\begin{center}
			{
				\setlength{\tabcolsep}{0mm}
				\begin{scriptsize}
					\begin{tabular*}{\textwidth}
						{
							@{}
							>{ \centering\arraybackslash}m{1.25cm} 
							>{ \centering\arraybackslash}m{1.32cm}				
							*{2}{ >{ \centering\arraybackslash}m{1.35cm}} 
							*{2}{ >{ \centering\arraybackslash}m{1.35cm}} 
							*{2}{ >{ \centering\arraybackslash}m{1.35cm}} 
							*{2}{ >{ \centering\arraybackslash}m{1.35cm}} 
							*{2}{ >{ \centering\arraybackslash}m{1.35cm}} 
							*{2}{ >{ \centering\arraybackslash}m{1.15cm}} 
							@{}
						}
						\multicolumn{1}{>{\centering\arraybackslash}m{1.25cm}}{\bf{XCAT Dataset}}&
						\multicolumn{1}{>{\centering\arraybackslash}m{1.32cm}}{\bf{Length (seconds)}}& 
						\multicolumn{2}{>{\centering\arraybackslash}m{2.7cm}}{\bf{3D tumor position reconstruction under 1 mm (\%)}}& 
						\multicolumn{2}{>{\centering\arraybackslash}m{2.7cm}}{\bf{3D tumor position reconstruction under 3 mm (\%)}}& 
						\multicolumn{2}{>{\centering\arraybackslash}m{2.7cm}}{\bf{3D tumor position reconstruction under 5 mm (\%)}}& 
						\multicolumn{2}{>{\centering\arraybackslash}m{2.7cm}}{\bf{3D tumor position reconstruction 95$^{\mathbf{th}}$P (mm)}}&
						\multicolumn{2}{>{\centering\arraybackslash}m{2.7cm}}{\bf{3D tumor position reconstruction RMSE (mm)}}&
						\multicolumn{2}{>{\centering\arraybackslash}m{2.3cm}}{\bf{3D image comparison NRMSE (\%)}}\\
						\hline
						& 		& FIR 	& COR  & FIR  & COR  & FIR  & COR  & FIR  & COR  & FIR  & COR  & FIR  & COR\\ 
						\hline
						1       & 61.8  & 59.9 & 36.6 & 94.2 & 83.2 & 96.8 & 93.5 & 3.21 & 6.31 & 1.46 & 2.64 & 98.29 & 97.96\\
						2       & 75.6  & 93.9 & 65.5 & 98.4 & 96.0 & 99.2 & 98.1 & 1.11 & 2.75 & 0.99 & 1.87 & 98.73 & 98.38\\
						3       & 46.7  & 75.1 & 76.4 & 100  & 99.1 & 100  & 100  & 1.51 & 2.22 & 0.85 & 0.96 & 96.87 & 95.96\\
						4       & 141.3 & 100  & 98.0 & 100  & 100  & 100  & 100  & 0.35 & 0.86 & 0.21 & 0.45 & 98.73 & 98.62\\
						5       & 161.4 & 95.7 & 54.6 & 100  & 97.6 & 100  & 99.8 & 0.97 & 2.62 & 0.54 & 1.36 & 98.22 & 97.65\\
						6       & 141.3 & 100  & 61.3 & 100  & 98.9 & 100  & 100  & 0.48 & 2.29 & 0.29 & 1.16 & 98.65 & 98.56\\
						7       & 169.1 & 49.9 & 45.3 & 94.7 & 100  & 100  & 100  & 2.02 & 2.39 & 1.16 & 1.39 & 98.39 & 98.37\\
						8       & 223.8 & 59.1 & 65.9 & 100  & 99.5 & 100  & 100  & 1.91 & 2.12 & 1.09 & 1.07 & 97.90 & 97.47\\
						9       & 245.0 & 13.5 & 26.8 & 97.4 & 73.8 & 100  & 100  & 2.93 & 3.90 & 2.13 & 2.40 & 97.97 & 97.79\\
						10      & 292.0 & 71.1 & 48.3 & 100  & 98.4 & 100  & 100  & 1.35 & 2.50 & 0.80 & 1.38 & 97.82 & 97.48\\
						\hline
						Average & 155.8 & 71.8 & 57.9 & 98.5 & 94.6 & 99.6 & 99.1 & 1.58 & 2.80 & 0.94 & 1.47 & 98.16 & 97.83\\
					\end{tabular*}
				\end{scriptsize}
			}
			\caption{Three dimensional tumor positions over an average respiratory trace duration of 155.8 seconds are reconstructed with a 95$^{\rm th}$ percentile of 1.58 mm with FIR, and 2.80 mm with the correlation model (COR) (averaged over 10 datasets). The RMSE between ground truth and FIR generated volumetric images is 0.94 mm with FIR, and the RMSE between ground truth and the correlation (COR) generated volumetric images is 1.47 mm (averaged over 10 datasets). The voxel by voxel intensity volumetric image comparison yields a normalized root mean squared error of 98.16\% with FIR, and a NRMSE of 97.83\% with the correlation model (COR) (averaged over 10 datasets).}
			\label{resultsRMSE}
		\end{center}
	\end{table*} 

	\section{Discussion}{\label{sec:Discussion}}
	This method provides a new way to continuously generate 3D images throughout a radiotherapy treatment by establishing a correlation model between motion model PCA weights determined from kV images (using information from the internal anatomy) and the amplitude and velocity of an external respiratory surrogate signal, which inherently takes into account any changes in the correlation between the internal and the external anatomy. This technique uses data that are often already available during treatments on conventional linear accelerators, and obviates the need to capture kV images at a high frame rate in order to accurately perform motion model-based 3D anatomical image reconstruction.
	
	Several techniques to track the tumor position under respiratory motion using fluoroscopic kV imaging have been developed \cite{KeallRealTimeIGART2018,KeallIGARTreview2018,ZhangIGART2018,NguyenKIMclinicalImplementation2017,KeallIGART2017,ColvillIGARTreview2016,KeallIGART2015,ChoIGART2011,XingIGARTchallenges2007}, however the additional dose due to the continuous kV image acquisition may be undesirable. The technique presented in this work enables anatomic tracking using only periodic kV imaging. Furthermore, by fitting a full motion model (rather than tracking individual points), the entire 3D patient anatomy can be reconstructed, enabling delivered dose calculation. 
	
	The correlation model used in this work differs from other techniques that have used internal-external correlation \cite{HeInternalExternalCorrelation2018,WuInternalExternalCorrelation2014,FayadInternalExternalCorrelation2011,LiuInternalExternalCorrelation2010,NishiokaInternalExternalCorrelation2008,BeddarInternalExternalCorrelation2007,KanoulasExternalSurrogate2007,SayehXLTS1p5mm2007,FuXLTS1p5mm2007,IonascuInternalExternalCorrelation2007,KorremanInternalExternalCorrelation2006,GiergaInternalExternalMarkersCorrelation2005} in that it relates both the surrogate amplitude and velocity to the PCA deformation vector fields in the motion model (as opposed to correlating surrogate amplitude and target position). Consequently this technique tracks both tumor positions and also determines the deformation of the entire patient's anatomy, including nearby organs at risk. The model can also be updated based on observed internal anatomy as more kV images are acquired. Additionally, as this technique uses information from the entire kV image (as opposed to just the tumor region), it does not suffer from limitations on tumor size or distance to other nearby structures that can often confound image-based tumor tracking techniques \cite{RubioSBRT2013}. As the {\it cine} 3D images produced by this method are generated from deformations of a reference 4DCT planning image, they are inherently co-registered with known deformations. The DVFs produced by the motion model can then be used to calculate and accumulate the delivered dose to the patient, providing a mechanism to compute the full delivered dose in the presence of respiratory motion during treatment.
	
	The 3D tumor positions are reconstructed with an average RMSE of 1.47 mm, which compares favorably with other similar tracking methods that use both images and external surrogates (e.g. the Xsight Lung Tracking System, which has a precision of 1.5 mm to 4 mm \cite{YangXLTS4mm2017,TorshabiFuzzyModel2013,HoogemanCyberKnifeClinicalAccuracy2009,SayehXLTS1p5mm2007,FuXLTS1p5mm2007,KilbyCyberKnife2010}). Performing volumetric image reconstruction with the FIR process using high frame rate kV images (0.2~second interval) yields an average RMSE of 0.94~mm in the tumor position reconstruction. Comparing to the results of the correlation model with a 3-second imaging interval shows that although the accuracy of the reconstruction decreases, the overall performance degradation (0.53~mm in the RMSE) is small compared to typical respiratory motion amplitudes. Voxelwise intensity comparison of both the FIR-generated and correlation model-generated volumetric images to the ground truth images yields a similar NRMSE, indicating good agreement between the methods in regions outside of the tumor.
	
	The use of XCAT phantoms enabled a detailed analysis of the full 3D accuracy of this technique. Although the phantoms used in this study match clinically observed respiratory patterns, the internal anatomic details of the phantom are not fully representative of a real patient. This could impact the performance of the deformable image registration as well as the fitting of the PCA weights from kV images. Additionally, physical effects such as scatter or imaging artifacts are not included in the image projection process. A full validation of this technique is currently undergoing with clinically acquired 4DCT and kV images. However, the ground truth comparison provided by this phantom study provides essential data to evaluate the accuracy of our technique and will be required to better understand results from the patient data analysis. As prior studies have already demonstrated the performance of kV-based PCA motion modeling techniques using clinical patient images, we believe the results of this analysis will generalize to clinically acquired imaging data.
	
	It is important to note that the XCAT phantoms used in this study are derived from tumor motion observed in patients using orthogonal imaging of implanted fiducial markers during radiotherapy treatments. Consequently, the respiratory motion patterns observed (rate, amplitude, variability) are true representations of the types of respiratory motions that can occur during clinical treatments. Furthermore, the external surrogate signal used for generating the correlation model was also recorded during treatment (not derived from the XCAT phantoms). This means that results presented in this work include any changes in internal-external correlation.
	
	The method presented in this work relies on a motion model built using DVFs generated from DIR between phases of a pre-treatment planning image set. Any uncertainties in these vector fields will be reflected in the motion model, and thus the accuracy of the DIR method places a limit on the precision of the generated images. Although a demons algorithm \cite{GuDemonsDIR2010} was used in this work, the method presented could be based on any DIR technique. Using registration techniques that may more accurately incorporate the biomechanical properties of respiratory motion may lead to better results \cite{SamavatiHybridDIR4DCT2015,BrockDIR2010}. Incorporating models developed from pre-treatment 4DCBCT \cite{Dhou4DCBCT2015} rather than a planning 4DCT may also improve the accuracy of the method by compensating for day-to-day changes in patient respiratory motion that may differ from the planning 4DCT model during treatment. Compensating for the changes in internal patient anatomy motion the occurs between the time of planning and the time of treatment could also be done by using subpopulation-based correspondence modeling \cite{WilmsImproved4DCT2017}. Moreover, some techniques have been proposed to model respiratory motion without 4DCT images obtained prior to treatment, which could further reduce the dose delivered to the patient in a clinical context \cite{FayadNo4DCTrequired2018,McClellandReconstructionFromPartialData2017}.
	
	This work uses a basic linear correlation model, and although the results (RMSE of 1.47 mm) are in the range expected for SBRT treatments, this technique could potentially be improved by developing a correlation model using a more advanced function or method, such as machine learning, neural networks \cite{SeregniArtificialNeuralNetworks2013,IsakssonAdaptiveNeuralNetwork2005}, nearest neighbor method \cite{ZhangNearestNeighborMethod2018} or fuzzy logic \cite{TorshabiFuzzyModel2013}. Moreover, the linear correlation model could be adjusted after every kV image acquisition \begin{hlbreakable}(at every 3 seconds)\end{hlbreakable} so that the continuous correlation model generated PCA weights are forced to agree with the periodic kV image derived PCA weights.
	
	\begin{hlbreakable}The maximum clinically available in-treatment imaging rate on treatment machines at our institution is set at 1 kV image every 3 seconds.  We used this maximum imaging frame rate in this study to assess the feasibility of this technique (and because we routinely use imaging intervals between 3-5 seconds to monitor motion in our clinical practice).  Changing the interval of kV images has the potential to impact the accuracy of the correlation model, and we plan to assess and optimize this rate in a future study.\end{hlbreakable} 
	
	In this study, the correlation model built for each dataset comes from the entire duration of the respiratory trace. However, the correlation between the internal and external anatomy can potentially change with time \cite{WuExternalSurrogate2008}, and implementing a time-adaptive correlation model that only uses a portion of the respiratory trace may improve accuracy. Additional improvements such as including the uncertainty in the FIR generation process within the model fitting will be investigated in future work.
	
	\begin{hlbreakable}The continuous volumetric images produced with this technique are inherently coregistered, as they are reconstructed by determining a DVF relative to a reference image. The high rate 3D anatomical image generation made possible with this technique allows to calculate accumulated delivered dose during the fraction with more precision, especially if the planning doses were calculated on the average intensity projection image, in which case discrepancies from the original treatment plan will be made apparent. Through this process, underdosage of the target or overdosage of critical structures can be identified, and corrected in subsequent fractions through treatment plan adaptation.\end{hlbreakable}

	\section{Conclusion}{\label{sec:Conclusion}}
	We have demonstrated a new technique for continuously generating dynamic volumetric images of patient anatomy using periodic kV imaging in combination with an external respiratory surrogate. The performance of the method was assessed with 10 digital XCAT phantoms that included clinically measured tumor positions and real patient breathing patterns, which enabled a comparison with ground truth tumor positions, demonstrating tumor localization to better than 1.47 mm on average. Correlation model generated volumetric images were reconstructed with a NRMSE of 97.83\% compared to the XCAT phantom ground truth volumetric images.
	
	This method is novel in its combination of kV image-based PCA motion modeling with an external respiratory surrogate amplitude and velocity for volumetric anatomical image reconstruction, which inherently include any changes in internal-external correlation. It enables motion modeling to be performed with high time resolution on data that can be acquired on current clinical linear accelerators (and in some cases is already routinely acquired in clinical practice), without additional kV imaging. The three-dimensional images produced using this technique can be used to calculate delivered dose in the presence of respiratory motion for thoracic and abdominal radiotherapy treatments. This will provide an important tool for treatment verification, delivered dose calculation, and adaptive radiotherapy.

	\section*{Acknowledgments}	
	We would like to thank Dr.~Nishioka of the Department of Radiology, NTT Hospital, Japan, Dr.~Shirato of the Department of Radiation Medicine, Hokkaido University, Japan, and Dr.~Berbeco of the Radiation Oncology Department at the Brigham \& Women's Hospital, Harvard Medical School, USA, for sharing the patient tumor position dataset with our group.     
	\newline
	
	This project was supported through a Master Research Agreement with Varian Medical Systems. 
	\newline
	\newline
	
	\center{\textbf{\textit{\Large{In memory of Guillaume Barlet.}}}}

	\bibliographystyle{elsarticle-num_noURL}
	\bibliography{Continuous_Generation_of_Volumetric_Images_Bibliography}

\begin{thebibliography}{100}
\expandafter\ifx\csname url\endcsname\relax
  \def\url#1{\texttt{#1}}\fi
\expandafter\ifx\csname urlprefix\endcsname\relax\def\urlprefix{URL }\fi
\expandafter\ifx\csname href\endcsname\relax
  \def\href#1#2{#2} \def\path#1{#1}\fi

\bibitem{AAPMTG76}
P.~Keall, G.~S. Mageras, J.~M. Balter, R.~S. Emery, K.~M. Forster, S.~B. Jiang,
  J.~M. Kapatoes, D.~A. Low, M.~J. Murphy, B.~R. Murray, C.~R. Ramsey, M.~B.
  van Herk, S.~S. Vedam, J.~W. Wong, E.~Yorke, The management of respiratory
  motion in radiation oncology report of aapm task group 76a), Medical Physics
  33~(10) (2006) 3874--3900
\newblock  (2006), \href {https://doi.org/10.1118/1.2349696}
  {\path{doi:10.1118/1.2349696}}.

\bibitem{KorremanIGRTmotionManagement2015}
S.~S. Korreman, Image-guided radiotherapy and motion management in lung cancer,
  The British Journal of Radiology 88~(1051) (2015) 20150100, pMID: 25955231
\newblock  (2015), \href {https://doi.org/10.1259/bjr.20150100}
  {\path{doi:10.1259/bjr.20150100}}.

\bibitem{KorremanMotionInRadiotherapy2012}
S.~S. Korreman, Motion in radiotherapy: photon therapy, Physics in Medicine and
  Biology 57~(23) (2012) R161--R191
\newblock  (nov 2012), \href {https://doi.org/10.1088/0031-9155/57/23/r161}
  {\path{doi:10.1088/0031-9155/57/23/r161}}.

\bibitem{JiangMobileTumorsRT2006}
S.~Jiang, Radiotherapy of mobile tumors, Seminars in Radiation Oncology 16~(4)
  (2006) 239 -- 248, innovative Technologies in Radiation Therapy
\newblock  (2006), \href
  {https://doi.org/https://doi.org/10.1016/j.semradonc.2006.04.007}
  {\path{doi:https://doi.org/10.1016/j.semradonc.2006.04.007}}.

\bibitem{Vedam4DCT2003}
S.~S. Vedam, P.~J. Keall, V.~R. Kini, H.~Mostafavi, H.~P. Shukla, R.~Mohan,
  Acquiring a four-dimensional computed tomography dataset using an external
  respiratory signal, Physics in Medicine \& Biology 48~(1) (2003) 45
\newblock  (2003).

\bibitem{GeoffreyTumorMotionEffectOnPlanning2003}
G.~D. Hugo, N.~Agazaryan, T.~D. Solberg, The effects of tumor motion on
  planning and delivery of respiratory-gated imrt, Medical Physics 30~(6)
  (2003) 1052--1066
\newblock  (2003), \href {https://doi.org/10.1118/1.1574611}
  {\path{doi:10.1118/1.1574611}}.

\bibitem{Rietzel4DCT2005}
E.~Rietzel, T.~Pan, G.~T.~Y. Chen, Four‐dimensional computed tomography:
  Image formation and clinical protocol, Medical Physics 32~(4) (2005) 874--889
\newblock  (2005), \href {https://doi.org/10.1118/1.1869852}
  {\path{doi:10.1118/1.1869852}}.

\bibitem{Pan4DCT2005}
T.~Pan, Comparison of helical and cine acquisitions for 4d-ct imaging with
  multislice ct, Medical Physics 32~(2) (2005) 627--634
\newblock  (2005), \href {https://doi.org/10.1118/1.1855013}
  {\path{doi:10.1118/1.1855013}}.

\bibitem{StJames4DCT2012}
S.~{St.~James}, P.~Mishra, F.~Hacker, R.~Berbeco, J.~Lewis, Quantifying itv
  instabilities arising from 4dct: a simulation study using patient data,
  Physics in Medicine \& Biology 57~(5) (2012) L1
\newblock  (2012).

\bibitem{Sarker4DCT2010}
J.~Sarker, A.~Chu, K.~Mui, J.~Wolfgang, A.~Hirsch, G.~Chen, G.~Sharp,
  Variations in tumor size and position due to irregular breathing in 4d‐ct:
  A simulation study, Medical Physics 37~(3) (2010) 1254--1260
\newblock  (2010), \href {https://doi.org/10.1118/1.3298007}
  {\path{doi:10.1118/1.3298007}}.

\bibitem{JingCaiGatedRT2010}
J.~Cai, R.~McLawhorn, P.~Read, J.~Larner, F.~Yin, S.~Benedict, K.~Sheng,
  Effects of breathing variation on gating window internal target volume in
  respiratory gated radiation therapy, Medical Physics 37~(8) (2010) 3927--3934
\newblock  (2010), \href {https://doi.org/10.1118/1.3457329}
  {\path{doi:10.1118/1.3457329}}.

\bibitem{JingCai4DCT2008}
J.~Cai, P.~Read, K.~Sheng, The effect of respiratory motion variability and
  tumor size on the accuracy of average intensity projection from
  four‐dimensional computed tomography: An investigation based on dynamic
  mri, Medical Physics 35~(11) (2008) 4974--4981
\newblock  (2008), \href {https://doi.org/10.1118/1.2982245}
  {\path{doi:10.1118/1.2982245}}.

\bibitem{StJames4DCT2016}
S.~{St.~James}, J.~Seco, P.~Mishra, J.~Lewis, Simulations using patient data to
  evaluate systematic errors that may occur in 4d treatment planning: A proof
  of concept study, Medical Physics 40~(9) (2016) 091706
\newblock  (2016), \href {https://doi.org/10.1118/1.4817244}
  {\path{doi:10.1118/1.4817244}}.

\bibitem{Mishra3DGeneration2013}
P.~Mishra, R.~Li, S.~{St.~James}, R.~Mak, C.~Williams, Y.~Yue, R.~Berbeco,
  J.~Lewis, Evaluation of 3d fluoroscopic image generation from a single planar
  treatment image on patient data with a modified xcat phantom., Physics in
  Medicine \& Biology 58~(4) (2013) 841
\newblock  (2013), \href {https://doi.org/10.1088/0031-9155/58/4/841}
  {\path{doi:10.1088/0031-9155/58/4/841}}.

\bibitem{ShiehBayesianMarkerless3DTracking2017}
C.-C. Shieh, V.~Caillet, M.~Dunbar, P.~J. Keall, J.~T. Booth, N.~Hardcastle,
  C.~Haddad, T.~Eade, I.~Feain, A bayesian approach for three-dimensional
  markerless tumor tracking using {kV} imaging during lung radiotherapy,
  Physics in Medicine and Biology 62~(8) (2017) 3065--3080
\newblock  (mar 2017), \href {https://doi.org/10.1088/1361-6560/aa6393}
  {\path{doi:10.1088/1361-6560/aa6393}}.

\bibitem{FassiExternalSurrogate2014}
A.~Fassi, J.~Schaerer, M.~Fernandes, M.~Riboldi, D.~Sarrut, G.~Baroni, Tumor
  tracking method based on a deformable 4d ct breathing motion model driven by
  an external surface surrogate, International Journal of Radiation
  Oncology*Biology*Physics 88~(1) (2014) 182 -- 188
\newblock  (2014), \href
  {https://doi.org/https://doi.org/10.1016/j.ijrobp.2013.09.026}
  {\path{doi:https://doi.org/10.1016/j.ijrobp.2013.09.026}}.

\bibitem{ZhangMotionArtifacts2010}
Q.~Zhang, Y.~Hu, F.~Liu, K.~Goodman, K.~Rosenzweig, G.~Mageras, Correction of
  motion artifacts in cone‐beam ct using a patient‐specific respiratory
  motion model, Medical Physics 37~(6Part1) (2010) 2901--2909
\newblock  (2010), \href {https://doi.org/10.1118/1.3397460}
  {\path{doi:10.1118/1.3397460}}.

\bibitem{LewisMarkerlessTracking2010}
J.~Lewis, R.~Li, W.~Watkins, J.~Lawson, P.~Segars, L.~Cervino, W.~Song,
  S.~Jiang, Markerless lung tumor tracking and trajectory reconstruction using
  rotational cone-beam projections: a feasibility study, Physics in Medicine \&
  Biology 55~(9) (2010) 2505
\newblock  (2010).

\bibitem{ZhangPatientSpecificMotionModel2007}
Q.~Zhang, A.~Pevsner, A.~Hertanto, Y.~Hu, K.~Rosenzweig, C.~Ling, G.~Mageras, A
  patient‐specific respiratory model of anatomical motion for radiation
  treatment planning, Medical Physics 34~(12) (2007) 4772--4781
\newblock  (2007), \href {https://doi.org/10.1118/1.2804576}
  {\path{doi:10.1118/1.2804576}}.

\bibitem{SohnPCA2005}
M.~Sohn, M.~Birkner, D.~Yan, M.~Alber, Modelling individual geometric variation
  based on dominant eigenmodes of organ deformation: implementation and
  evaluation, Physics in Medicine \& Biology 50~(24) (2005) 5893
\newblock  (2005).

\bibitem{HertantoMotionModel2012}
A.~Hertanto, Q.~Zhang, Y.~Hu, O.~Dzyubak, A.~Rimner, G.~Mageras, Reduction of
  irregular breathing artifacts in respiration‐correlated ct images using a
  respiratory motion model, Medical Physics 39~(6Part1) (2012) 3070--3079
\newblock  (2012), \href {https://doi.org/10.1118/1.4711802}
  {\path{doi:10.1118/1.4711802}}.

\bibitem{LiPCA2011}
R.~Li, J.~Lewis, X.~Jia, T.~Zhao, W.~Liu, S.~Wuenschel, J.~Lamb, D.~Yang,
  D.~Low, S.~Jiang, On a pca-based lung motion model, Physics in Medicine \&
  Biology 56~(18) (2011) 6009
\newblock  (2011).

\bibitem{LowMotionModel2005}
D.~A. Low, P.~J. Parikh, W.~Lu, J.~F. Dempsey, S.~H. Wahab, J.~P. Hubenschmidt,
  M.~M. Nystrom, M.~Handoko, J.~D. Bradley, Novel breathing motion model for
  radiotherapy, International Journal of Radiation Oncology*Biology*Physics
  63~(3) (2005) 921 -- 929
\newblock  (2005), \href
  {https://doi.org/https://doi.org/10.1016/j.ijrobp.2005.03.070}
  {\path{doi:https://doi.org/10.1016/j.ijrobp.2005.03.070}}.

\bibitem{CaiDoseAssessment2015}
W.~Cai, M.~Hurwitz, C.~Williams, S.~Dhou, R.~Berbeco, J.~Seco, P.~Mishra,
  J.~Lewis, 3d delivered dose assessment using a 4dct‐based motion model,
  Medical Physics 42~(6Part1) (2015) 2897--2907
\newblock  (2015), \href {https://doi.org/10.1118/1.4921041}
  {\path{doi:10.1118/1.4921041}}.

\bibitem{Dhou4DCBCT2015}
S.~Dhou, M.~Hurwitz, P.~Mishra, W.~Cai, J.~Rottmann, R.~Li, C.~Williams,
  M.~Wagar, R.~Berbeco, D.~Ionascu, J.~Lewis, 3d fluoroscopic image estimation
  using patient-specific 4dcbct-based motion models, Physics in Medicine \&
  Biology 60~(9) (2015) 3807
\newblock  (2015).

\bibitem{HurwitzExternalSurrogate2015}
M.~Hurwitz, C.~Williams, P.~Mishra, J.~Rottmann, S.~Dhou, M.~Wagar,
  E.~Mannarino, R.~Mak, J.~Lewis, Generation of fluoroscopic 3d images with a
  respiratory motion model based on an external surrogate signal, Physics in
  Medicine \& Biology 60~(2) (2015) 521
\newblock  (2015).

\bibitem{McClellandRespiratorySurrogate2011}
J.~McClelland, S.~Hughes, M.~Modat, A.~Qureshi, S.~Ahmad, D.~Landau,
  S.~Ourselin, D.~Hawkes, Inter-fraction variations in respiratory motion
  models, Physics in Medicine \& Biology 56~(1) (2011) 251
\newblock  (2011).

\bibitem{Staub4DCBCTPCA2011}
D.~Staub, A.~Docef, R.~Brock, C.~Vaman, M.~Murphy, 4d cone‐beam ct
  reconstruction using a motion model based on principal component analysis,
  Medical Physics 38~(12) (2011) 6697--6709
\newblock  (2011), \href {https://doi.org/10.1118/1.3662895}
  {\path{doi:10.1118/1.3662895}}.

\bibitem{ChoXrayTumorTrackingWithSurrogate2008}
B.~Cho, Y.~Suh, S.~Dieterich, P.~Keall, A monoscopic method for real-time
  tumour tracking using combined occasional x-ray imaging and continuous
  respiratory monitoring, Physics in Medicine \& Biology 53~(11) (2008) 2837
\newblock  (2008).

\bibitem{HarrisKVimagesTo4DMRI2018}
W.~Harris, C.~Wang, F.~Yin, J.~Cai, L.~Ren, A novel method to generate on-board
  4d mri using prior 4d mri and on-board kv projections from a conventional
  linac for target localization in liver sbrt, Medical Physics 45~(7) (2018)
  3238--3245
\newblock  (2018), \href {https://doi.org/10.1002/mp.12998}
  {\path{doi:10.1002/mp.12998}}.

\bibitem{MishraMVimages2014}
P.~Mishra, R.~Li, R.~Mak, J.~Rottmann, J.~Bryant, C.~Williams, R.~Berbeco,
  J.~Lewis, An initial study on the estimation of time‐varying volumetric
  treatment images and 3d tumor localization from single mv cine epid images,
  Medical Physics 41~(8Part1) (2014) 081713
\newblock  (2014), \href {https://doi.org/10.1118/1.4889779}
  {\path{doi:10.1118/1.4889779}}.

\bibitem{LiSurrogateBasedVMAT2012}
R.~Li, E.~Mok, B.~Han, A.~Koong, L.~Xing, Evaluation of the geometric accuracy
  of surrogate-based gated vmat using intrafraction kilovoltage x-ray images,
  Medical Physics 39~(5) (2012) 2686--2693
\newblock  (2012), \href {https://doi.org/10.1118/1.4704729}
  {\path{doi:10.1118/1.4704729}}.

\bibitem{NgContinuousKVimagingDose2013}
J.~A. Ng, J.~Booth, P.~Poulsen, Z.~Kuncic, P.~J. Keall, Estimation of effective
  imaging dose for kilovoltage intratreatment monitoring of the prostate
  position during cancer radiotherapy, Physics in Medicine and Biology 58~(17)
  (2013) 5983--5996
\newblock  (aug 2013), \href {https://doi.org/10.1088/0031-9155/58/17/5983}
  {\path{doi:10.1088/0031-9155/58/17/5983}}.

\bibitem{CrockerContinuousKVimagingDose2012}
J.~K. Crocker, J.~A. Ng, P.~J. Keall, J.~T. Booth, Measurement of patient
  imaging dose for real-time kilovoltage x-ray intrafraction tumour position
  monitoring in prostate patients, Physics in Medicine and Biology 57~(10)
  (2012) 2969--2980
\newblock  (apr 2012), \href {https://doi.org/10.1088/0031-9155/57/10/2969}
  {\path{doi:10.1088/0031-9155/57/10/2969}}.

\bibitem{KeallIGART2015}
P.~J. Keall, J.~Aun~Ng, R.~O'Brien, E.~Colvill, C.-Y. Huang,
  P.~Rugaard~Poulsen, W.~Fledelius, P.~Juneja, E.~Simpson, L.~Bell, F.~Alfieri,
  T.~Eade, A.~Kneebone, J.~T. Booth, The first clinical treatment with
  kilovoltage intrafraction monitoring (kim): A real-time image guidance
  method, Medical Physics 42~(1) (2015) 354--358
\newblock  (2015), \href {https://doi.org/10.1118/1.4904023}
  {\path{doi:10.1118/1.4904023}}.

\bibitem{WuExternalSurrogate2008}
H.~Wu, Q.~Zhao, R.~Berbeco, S.~Nishioka, H.~Shirato, S.~Jiang, Gating based on
  internal/external signals with dynamic correlation updates, Physics in
  Medicine \& Biology 53~(24) (2008) 7137
\newblock  (2008).

\bibitem{KanoulasExternalSurrogate2007}
E.~Kanoulas, J.~Aslam, G.~Sharp, R.~Berbeco, S.~Nishioka, H.~Shirato, S.~Jiang,
  Derivation of the tumor position from external respiratory surrogates with
  periodical updating of the internal/external correlation, Physics in Medicine
  \& Biology 52~(17) (2007) 5443
\newblock  (2007).

\bibitem{BerbecoExternalSurrogate2005}
R.~I. Berbeco, S.~Nishioka, H.~Shirato, G.~T.~Y. Chen, S.~B. Jiang, Residual
  motion of lung tumours in gated radiotherapy with external respiratory
  surrogates, Physics in Medicine \& Biology 50~(16) (2005) 3655
\newblock  (2005).

\bibitem{AdlerOrthogonalXRayAndExternalSurfaceMonitoringCyberknife2000}
A.~Schweikard, G.~Glosser, M.~Bodduluri, M.~J. Murphy, J.~R. Adler, Robotic
  motion compensation for respiratory movement during radiosurgery, Computer
  Aided Surgery 5~(4) (2000) 263--277, pMID: 11029159
\newblock  (2000), \href {https://doi.org/10.3109/10929080009148894}
  {\path{doi:10.3109/10929080009148894}}.

\bibitem{YangXLTS4mm2017}
Z.-Y. Yang, Y.~Chang, H.-Y. Liu, G.~Liu, Q.~Li, Target margin design for
  real-time lung tumor tracking stereotactic body radiation therapy using
  cyberknife xsight lung tracking system, in: Scientific Reports, 2017, pp.
  1--11
\newblock  (2017), \href {https://doi.org/10.1038/s41598-017-11128-w}
  {\path{doi:10.1038/s41598-017-11128-w}}.

\bibitem{TorshabiFuzzyModel2013}
A.~Torshabi, M.~Riboldi, A.~Fooladi, S.~Mosalla, G.~Baroni, An adaptive fuzzy
  prediction model for real time tumor tracking in radiotherapy via external
  surrogates, Journal of Applied Clinical Medical Physics 14~(1) (2013)
  102--114
\newblock  (2013), \href {https://doi.org/10.1120/jacmp.v14i1.4008}
  {\path{doi:10.1120/jacmp.v14i1.4008}}.

\bibitem{CollinsCyberknifeRealTimeTumorMotionTracking2007}
B.~T. Collins, K.~Erickson, C.~A. Reichner, S.~P. Collins, G.~J. Gagnon,
  S.~Dieterich, D.~A. McRae, Y.~Zhang, S.~Yousefi, E.~Levy, T.~Chang,
  C.~Jamis-Dow, F.~Banovac, E.~D. Anderson, Radical stereotactic radiosurgery
  with real-time tumor motion tracking in the treatment of small peripheral
  lung tumors, Radiation Oncology 2~(1) (2007) 39
\newblock  (Oct 2007), \href {https://doi.org/10.1186/1748-717X-2-39}
  {\path{doi:10.1186/1748-717X-2-39}}.

\bibitem{SayehXLTS1p5mm2007}
S.~Sayeh, J.~Wang, W.~Main, W.~Kilby, C.~Maurer, Respiratory Motion Tracking
  for Robotic Radiosurgery, Springer, 2007
\newblock  (01 2007).

\bibitem{FuXLTS1p5mm2007}
D.~Fu, R.~Kahn, B.~Wang, H.~Wang, Z.~Mu, J.~Park, G.~Kuduvalli, C.~Maurer,
  Xsight Lung Tracking System: A Fiducial-Less Method for Respiratory Motion
  Tracking, Springer, 2007
\newblock  (01 2007).

\bibitem{KilbyCyberKnife2010}
W.~Kilby, J.~Dooley, G.~Kuduvalli, S.~Sayeh, C.~{Maurer}, The cyberknife
  robotic radiosurgery system in 2010, Technology in Cancer Research \&
  Treatment 9~(5) (2010) 433--452, pMID: 20815415
\newblock  (2010), \href {https://doi.org/10.1177/153303461000900502}
  {\path{doi:10.1177/153303461000900502}}.

\bibitem{KuboBSRT2000}
H.~D. Kubo, P.~M. Len, S.-i. Minohara, H.~Mostafavi, Breathing-synchronized
  radiotherapy program at the university of california davis cancer center,
  Medical Physics 27~(2) (2000) 346--353
\newblock  (2000), \href {https://doi.org/10.1118/1.598837}
  {\path{doi:10.1118/1.598837}}.

\bibitem{FordRPM2002522}
E.~Ford, G.~Mageras, E.~Yorke, K.~Rosenzweig, R.~Wagman, C.~Ling, Evaluation of
  respiratory movement during gated radiotherapy using film and electronic
  portal imaging, International Journal of Radiation Oncology*Biology*Physics
  52~(2) (2002) 522 -- 531
\newblock  (2002), \href
  {https://doi.org/https://doi.org/10.1016/S0360-3016(01)02681-5}
  {\path{doi:https://doi.org/10.1016/S0360-3016(01)02681-5}}.

\bibitem{WagmanRPM2003659}
R.~Wagman, E.~Yorke, E.~Ford, P.~Giraud, G.~Mageras, B.~Minsky, K.~Rosenzweig,
  Respiratory gating for liver tumors: use in dose escalation, International
  Journal of Radiation Oncology*Biology*Physics 55~(3) (2003) 659 -- 668
\newblock  (2003), \href
  {https://doi.org/https://doi.org/10.1016/S0360-3016(02)03941-X}
  {\path{doi:https://doi.org/10.1016/S0360-3016(02)03941-X}}.

\bibitem{KeallRPM200481}
P.~Keall, 4-dimensional computed tomography imaging and treatment planning,
  Seminars in Radiation Oncology 14~(1) (2004) 81 -- 90, high-Precision
  Radiation Therapy of Moving Targets
\newblock  (2004), \href
  {https://doi.org/https://doi.org/10.1053/j.semradonc.2003.10.006}
  {\path{doi:https://doi.org/10.1053/j.semradonc.2003.10.006}}.

\bibitem{GiraudRespiratoryGatedRadiotherapy2005}
P.~Giraud, L.~Simon, M.~Saliou, F.~Reboul, R.~Garcia, C.~Carrie, U.~Lerolle,
  J.~C. Rosenwald, J.~M. Cosset, {[{R}espiratory gated radiotherapy: the 4{D}
  radiotherapy]}, Bull Cancer 92~(1) (2005) 83--89
\newblock  (Jan 2005).

\bibitem{GiraudRespiratoryGating2010}
P.~Giraud, R.~Garcia, {[{R}espiratory gating for radiotherapy: main technical
  aspects and clinical benefits]}, Bull Cancer 97~(7) (2010) 847--856
\newblock  (Jul 2010).

\bibitem{LowSpirometry4DCT2003}
D.~A. Low, M.~Nystrom, E.~Kalinin, P.~Parikh, J.~F. Dempsey, J.~D. Bradley,
  S.~Mutic, S.~H. Wahab, T.~Islam, G.~Christensen, D.~G. Politte, B.~R.
  Whiting, A method for the reconstruction of four-dimensional synchronized ct
  scans acquired during free breathing, Medical Physics 30~(6) (2003)
  1254--1263
\newblock  (2003), \href {https://doi.org/10.1118/1.1576230}
  {\path{doi:10.1118/1.1576230}}.

\bibitem{ZhangSpirometerBasedPlanOptimization2004}
T.~Zhang, R.~Jeraj, H.~Keller, W.~Lu, G.~H. Olivera, T.~R. McNutt, T.~R.
  Mackie, B.~Paliwal, Treatment plan optimization incorporating respiratory
  motion, Medical Physics 31~(6) (2004) 1576--1586
\newblock  (2004), \href {https://doi.org/10.1118/1.1739672}
  {\path{doi:10.1118/1.1739672}}.

\bibitem{ZhangSpirometerBasedDIBH2003}
T.~Zhang, H.~Keller, M.~J. O'Brien, T.~R. Mackie, B.~Paliwal, Application of
  the spirometer in respiratory gated radiotherapy, Medical Physics 30~(12)
  (2003) 3165--3171
\newblock  (2003), \href {https://doi.org/10.1118/1.1625439}
  {\path{doi:10.1118/1.1625439}}.

\bibitem{ZhangSpirometryPatientCoaching2003}
T.~Zhang, H.~Keller, R.~Jeraj, R.~Manon, J.~Welsh, R.~Patel, J.~Fenwick,
  M.~Mehta, T.~Mackie, B.~Paliwal, Breathing synchronized delivery - a new
  technique for radiation treatment of the targets with respiratory motion,
  International Journal of Radiation Oncology*Biology*Physics 57~(2,
  Supplement) (2003) S185 -- S186
\newblock  (2003), \href
  {https://doi.org/https://doi.org/10.1016/S0360-3016(03)00980-5}
  {\path{doi:https://doi.org/10.1016/S0360-3016(03)00980-5}}.

\bibitem{SegarsXCAT2010}
P.~Segars, G.~Sturgeon, S.~Mendonca, J.~Grimes, B.~Tsui, 4d xcat phantom for
  multimodality imaging research, Medical Physics 37~(9) (2010) 4902--4915
\newblock  (2010), \href {https://doi.org/10.1118/1.3480985}
  {\path{doi:10.1118/1.3480985}}.

\bibitem{SegarsThesis}
P.~Segars, B.~Tsui, D.~Lalush, E.~Frey, M.~King, D.~Manocha, Development and
  application of the new dynamic nurbs-based cardiac-torso (ncat) phantom.,
  Journal OF Nuclear Medecine 42 (2001) 23
\newblock  (05 2001).

\bibitem{SegarsMCAT2001}
P.~Segars, D.~Lalush, B.~Tsui, Modeling respiratory mechanics in the mcat and
  spline-based mcat phantoms, IEEE Transactions on Nuclear Science 48~(1)
  (2001) 89--97
\newblock  (02 2001), \href {https://doi.org/10.1109/23.910837}
  {\path{doi:10.1109/23.910837}}.

\bibitem{SpitzerVisibleHumanDataset1998}
V.~Spitzer, D.~Whitlock, The visible human dataset: The anatomical platform for
  human simulation, The Anatomical Record 253~(2) (1998) 49--57
\newblock  (1998), \href
  {https://doi.org/10.1002/(SICI)1097-0185(199804)253:2<49::AID-AR8>3.0.CO;2-9}
  {\path{doi:10.1002/(SICI)1097-0185(199804)253:2<49::AID-AR8>3.0.CO;2-9}}.

\bibitem{BerbecoFluoroscopicGating2005}
R.~I. Berbeco, H.~Mostafavi, G.~C. Sharp, S.~B. Jiang, Towards fluoroscopic
  respiratory gating for lung tumours without radiopaque markers, Physics in
  Medicine \& Biology 50~(19) (2005) 4481
\newblock  (2005).

\bibitem{MishraModifiedXCAT2012}
M.~Pankaj, S.~{St.~James}, P.~Segars, R.~Berbeco, J.~Lewis, Adaptation and
  applications of a realistic digital phantom based on patient lung tumor
  trajectories, Physics in Medicine \& Biology 57~(11) (2012) 3597
\newblock  (2012).

\bibitem{CaiDoseAssessmentSBRT2015}
W.~Cai, S.~Dhou, F.~Cifter, M.~Myronakis, M.~Hurwitz, C.~Williams, R.~Berbeco,
  J.~Seco, J.~Lewis, 4d cone beam ct-based dose assessment for sbrt lung cancer
  treatment, Physics in Medicine \& Biology 61~(2) (2016) 554
\newblock  (2016).

\bibitem{Williams4DXCAT2013}
C.~Williams, P.~Mishra, J.~Seco, S.~{St.~James}, R.~Mak, R.~Berbeco, J.~Lewis,
  A mass-conserving 4d xcat phantom for dose calculation and accumulation,
  Medical Physics 40~(7) (2013) 071728
\newblock  (2013), \href {https://doi.org/10.1118/1.4811102}
  {\path{doi:10.1118/1.4811102}}.

\bibitem{ShiratoTreatmentPlanning2000}
H.~Shirato, S.~Shimizu, K.~Kitamura, T.~Nishioka, K.~Kagei, S.~Hashimoto,
  H.~Aoyama, T.~Kunieda, N.~Shinohara, H.~Dosaka-Akita, K.~Miyasaka,
  Four-dimensional treatment planning and fluoroscopic real-time tumor tracking
  radiotherapy for moving tumor, International Journal of Radiation
  Oncology*Biology*Physics 48~(2) (2000) 435 -- 442
\newblock  (2000), \href
  {https://doi.org/https://doi.org/10.1016/S0360-3016(00)00625-8}
  {\path{doi:https://doi.org/10.1016/S0360-3016(00)00625-8}}.

\bibitem{ShiratoHokkaidoDataset2000}
H.~Shirato, S.~Shimizu, T.~Kunieda, K.~Kitamura, M.~van Herk, K.~Kagei,
  T.~Nishioka, S.~Hashimoto, K.~Fujita, H.~Aoyama, K.~Tsuchiya, K.~Kudo,
  K.~Miyasaka, Physical aspects of a real-time tumor-tracking system for gated
  radiotherapy, International Journal of Radiation Oncology*Biology*Physics
  48~(4) (2000) 1187 -- 1195
\newblock  (2000), \href
  {https://doi.org/https://doi.org/10.1016/S0360-3016(00)00748-3}
  {\path{doi:https://doi.org/10.1016/S0360-3016(00)00748-3}}.

\bibitem{BerbecoAnzai733V2010}
R.~I. Berbeco, S.~Nishioka, H.~Shirato, Evaluation of the need for simultaneous
  orthogonal gated setup imaging, Journal of Applied Clinical Medical Physics
  11~(2) (2010) 158--167
\newblock  (2010), \href {https://doi.org/10.1120/jacmp.v11i2.3203}
  {\path{doi:10.1120/jacmp.v11i2.3203}}.

\bibitem{MyronakisXCATinterface2017}
M.~Myronakis, W.~Cai, S.~Dhou, F.~Cifter, M.~Hurwitz, P.~Segars, R.~Berbeco,
  J.~Lewis, A graphical user interface for xcat phantom configuration,
  generation and processing, Biomedical Physics \& Engineering Express 3~(1)
  (2017) 017003
\newblock  (2017).

\bibitem{Li3DtumorLoc2011}
R.~Li, J.~Lewis, X.~Jia, X.~Gu, M.~Folkerts, C.~Men, W.~Song, S.~Jiang, 3d
  tumor localization through real‐time volumetric x‐ray imaging for lung
  cancer radiotherapy, Medical Physics 38~(5) (2011) 2783--2794
\newblock  (2011), \href {https://doi.org/10.1118/1.3582693}
  {\path{doi:10.1118/1.3582693}}.

\bibitem{GuDemonsDIR2010}
X.~Gu, H.~Pan, Y.~Liang, R.~Castillo, D.~Yang, D.~Choi, E.~Castillo,
  A.~Majumdar, T.~Guerrero, S.~Jiang, Implementation and evaluation of various
  demons deformable image registration algorithms on a gpu, Physics in Medicine
  \& Biology 55~(1) (2010) 207
\newblock  (2010).

\bibitem{ZhangPCA2013}
Y.~Zhang, J.~Yang, L.~Zhang, L.~Court, P.~Balter, L.~Dong, Modeling respiratory
  motion for reducing motion artifacts in 4d ct images, Medical Physics 40~(4)
  (2013) 041716
\newblock  (2013), \href {https://doi.org/10.1118/1.4795133}
  {\path{doi:10.1118/1.4795133}}.

\bibitem{LiVolumetricReconstruction2010}
R.~Li, X.~Jia, J.~Lewis, X.~Gu, M.~Folkerts, C.~Men, S.~Jiang, Real‐time
  volumetric image reconstruction and 3d tumor localization based on a single
  x‐ray projection image for lung cancer radiotherapy, Medical Physics
  37~(6Part1) (2010) 2822--2826
\newblock  (2010), \href {https://doi.org/10.1118/1.3426002}
  {\path{doi:10.1118/1.3426002}}.

\bibitem{Vaman4DCTregistration2010}
C.~Vaman, D.~Staub, J.~Williamson, M.~Murphy, A method to map errors in the
  deformable registration of 4dct images, Medical Physics 37~(11) (2010)
  5765--5776
\newblock  (2010), \href {https://doi.org/10.1118/1.3488983}
  {\path{doi:10.1118/1.3488983}}.

\bibitem{ZengTomographicRegistration2007}
R.~Zeng, J.~A. Fessler, J.~Balter, Estimating 3-d respiratory motion from
  orbiting views by tomographic image registration, IEEE Transactions on
  Medical Imaging 26~(2) (2007) 153--163
\newblock  (02 2007), \href {https://doi.org/10.1109/TMI.2006.889719}
  {\path{doi:10.1109/TMI.2006.889719}}.

\bibitem{Zhao5DlungMotionModel2009}
T.~Zhao, W.~Lu, D.~Yang, S.~Mutic, C.~Noel, P.~Parikh, J.~Bradley, D.~Low,
  Characterization of free breathing patterns with 5d lung motion model,
  Medical Physics 36~(11) (2009) 5183--5189
\newblock  (2009), \href {https://doi.org/10.1118/1.3246348}
  {\path{doi:10.1118/1.3246348}}.

\bibitem{KeallRealTimeIGART2018}
P.~J. Keall, D.~T. Nguyen, R.~O'Brien, V.~Caillet, E.~Hewson, P.~R. Poulsen,
  R.~Bromley, L.~Bell, T.~Eade, A.~Kneebone, J.~Martin, J.~T. Booth, The first
  clinical implementation of real-time image-guided adaptive radiotherapy using
  a standard linear accelerator, Radiotherapy and Oncology 127~(1) (2018) 6 --
  11
\newblock  (2018), \href
  {https://doi.org/https://doi.org/10.1016/j.radonc.2018.01.001}
  {\path{doi:https://doi.org/10.1016/j.radonc.2018.01.001}}.

\bibitem{KeallIGARTreview2018}
P.~J. Keall, D.~T. Nguyen, R.~O'Brien, P.~Zhang, L.~Happersett, J.~Bertholet,
  P.~R. Poulsen, Review of real-time 3-dimensional image guided radiation
  therapy on standard-equipped cancer radiation therapy systems: Are we at the
  tipping point for the era of real-time radiation therapy?, International
  Journal of Radiation Oncology*Biology*Physics 102~(4) (2018) 922 -- 931,
  imaging in Radiation Oncology
\newblock  (2018), \href
  {https://doi.org/https://doi.org/10.1016/j.ijrobp.2018.04.016}
  {\path{doi:https://doi.org/10.1016/j.ijrobp.2018.04.016}}.

\bibitem{ZhangIGART2018}
P.~Zhang, M.~Hunt, A.~B. Telles, H.~Pham, M.~Lovelock, E.~Yorke, G.~Li,
  L.~Happersett, A.~Rimner, G.~Mageras, Design and validation of a mv/kv
  imaging-based markerless tracking system for assessing real-time lung tumor
  motion, Medical Physics 45~(12) (2018) 5555--5563
\newblock  (2018), \href {https://doi.org/10.1002/mp.13259}
  {\path{doi:10.1002/mp.13259}}.

\bibitem{NguyenKIMclinicalImplementation2017}
D.~T. Nguyen, R.~O'Brien, J.-H. Kim, C.-Y. Huang, L.~Wilton, P.~Greer,
  K.~Legge, J.~T. Booth, P.~R. Poulsen, J.~Martin, P.~J. Keall, The first
  clinical implementation of a real-time six degree of freedom target tracking
  system during radiation therapy based on kilovoltage intrafraction monitoring
  (kim), Radiotherapy and Oncology 123~(1) (2017) 37 -- 42
\newblock  (2017), \href
  {https://doi.org/https://doi.org/10.1016/j.radonc.2017.02.013}
  {\path{doi:https://doi.org/10.1016/j.radonc.2017.02.013}}.

\bibitem{KeallIGART2017}
P.~Keall, D.~T. Nguyen, R.~O'Brien, J.~Booth, P.~Greer, P.~Poulsen, V.~Gebski,
  A.~Kneebone, J.~Martin, Stereotactic prostate adaptive radiotherapy utilising
  kilovoltage intrafraction monitoring: the trog 15.01 spark trial, BMC Cancer
  17~(1) (2017) 180
\newblock  (Mar 2017), \href {https://doi.org/10.1186/s12885-017-3164-1}
  {\path{doi:10.1186/s12885-017-3164-1}}.

\bibitem{ColvillIGARTreview2016}
E.~Colvill, J.~Booth, S.~Nill, M.~Fast, J.~Bedford, U.~Oelfke, M.~Nakamura,
  P.~Poulsen, E.~Worm, R.~Hansen, T.~Ravkilde, J.~S. Rydhög, T.~Pommer, P.~M.
  af~Rosenschold, S.~Lang, M.~Guckenberger, C.~Groh, C.~Herrmann, D.~Verellen,
  K.~Poels, L.~Wang, M.~Hadsell, T.~Sothmann, O.~Blanck, P.~Keall, A dosimetric
  comparison of real-time adaptive and non-adaptive radiotherapy: A
  multi-institutional study encompassing robotic, gimbaled, multileaf
  collimator and couch tracking, Radiotherapy and Oncology 119~(1) (2016) 159
  -- 165
\newblock  (2016), \href
  {https://doi.org/https://doi.org/10.1016/j.radonc.2016.03.006}
  {\path{doi:https://doi.org/10.1016/j.radonc.2016.03.006}}.

\bibitem{ChoIGART2011}
B.~Cho, P.~R. Poulsen, A.~Sawant, D.~Ruan, P.~J. Keall, Real-time target
  position estimation using stereoscopic kilovoltage/megavoltage imaging and
  external respiratory monitoring for dynamic multileaf collimator tracking,
  International Journal of Radiation Oncology*Biology*Physics 79~(1) (2011) 269
  -- 278
\newblock  (2011), \href
  {https://doi.org/https://doi.org/10.1016/j.ijrobp.2010.02.052}
  {\path{doi:https://doi.org/10.1016/j.ijrobp.2010.02.052}}.

\bibitem{XingIGARTchallenges2007}
L.~Xing, J.~Siebers, P.~Keall, Computational challenges for image-guided
  radiation therapy: Framework and current research, Seminars in Radiation
  Oncology 17~(4) (2007) 245 -- 257, image-Guided Radiation Therapy
\newblock  (2007), \href
  {https://doi.org/https://doi.org/10.1016/j.semradonc.2007.07.004}
  {\path{doi:https://doi.org/10.1016/j.semradonc.2007.07.004}}.

\bibitem{HeInternalExternalCorrelation2018}
P.~He, Q.~Li, G.~Xiao, X.~Wang, S.~Ouyang, R.~Liu, Effect of respiratory
  guidance on internal/external respiratory motion correlation for
  synchrotron-based pulsed heavy-ion radiotherapy, Australasian Physical {\&}
  Engineering Sciences in Medicine 41~(3) (2018) 713--720
\newblock  (09 2018), \href {https://doi.org/10.1007/s13246-018-0667-2}
  {\path{doi:10.1007/s13246-018-0667-2}}.

\bibitem{WuInternalExternalCorrelation2014}
H.~Wu, A.~Besemer, M.~Lu, Dynamic correlation of synchronized internal tumor
  and external skin respiratory motion for image guided lung cancer radiation
  treatment, International Journal of Computers and their Applications 21~(1)
  (2014) 14--23
\newblock  (3 2014).

\bibitem{FayadInternalExternalCorrelation2011}
H.~Fayad, T.~Pan, J.~Clement, D.~Visvikis, Technical note: Correlation of
  respiratory motion between external patient surface and internal anatomical
  landmarks, Medical Physics 38~(6Part1) (2011) 3157--3164
\newblock  (2011), \href {https://doi.org/10.1118/1.3589131}
  {\path{doi:10.1118/1.3589131}}.

\bibitem{LiuInternalExternalCorrelation2010}
C.~Liu, A.~Alessio, P.~Kinahan, Respiratory motion correction for pet/ct with
  internal-external motion correlation, Journal of Nuclear Medicine
  51~(supplement 2) (2010) 78
\newblock  (2010).

\bibitem{NishiokaInternalExternalCorrelation2008}
S.~Nishioka, T.~Nishioka, M.~Kawahara, S.~Tanaka, T.~Hiromura, K.~Tomita,
  H.~Shirato, Exhale fluctuation in respiratory-gated radiotherapy of the lung:
  A pitfall of respiratory gating shown in a synchronized internal/external
  marker recording study, Radiotherapy and Oncology 86~(1) (2008) 69 -- 76
\newblock  (2008), \href
  {https://doi.org/https://doi.org/10.1016/j.radonc.2007.11.014}
  {\path{doi:https://doi.org/10.1016/j.radonc.2007.11.014}}.

\bibitem{BeddarInternalExternalCorrelation2007}
A.~S. Beddar, K.~Kainz, T.~M. Briere, Y.~Tsunashima, T.~Pan, K.~Prado,
  R.~Mohan, M.~Gillin, S.~Krishnan, Correlation between internal fiducial tumor
  motion and external marker motion for liver tumors imaged with 4d-ct,
  International Journal of Radiation Oncology*Biology*Physics 67~(2) (2007) 630
  -- 638
\newblock  (2007), \href
  {https://doi.org/https://doi.org/10.1016/j.ijrobp.2006.10.007}
  {\path{doi:https://doi.org/10.1016/j.ijrobp.2006.10.007}}.

\bibitem{IonascuInternalExternalCorrelation2007}
D.~Ionascu, S.~Jiang, S.~Nishioka, H.~Shirato, R.~Berbeco, Internal-external
  correlation investigations of respiratory induced motion of lung tumors,
  Medical Physics 34~(10) (2007) 3893--3903
\newblock  (2007), \href {https://doi.org/10.1118/1.2779941}
  {\path{doi:10.1118/1.2779941}}.

\bibitem{KorremanInternalExternalCorrelation2006}
S.~Korreman, H.~Mostafavi, Q.-T. Le, A.~Boyer, Comparison of respiratory
  surrogates for gated lung radiotherapy without internal fiducials, Acta
  Oncologica 45~(7) (2006) 935--942
\newblock  (2006), \href {https://doi.org/10.1080/02841860600917161}
  {\path{doi:10.1080/02841860600917161}}.

\bibitem{GiergaInternalExternalMarkersCorrelation2005}
D.~Gierga, J.~Brewer, G.~Sharp, M.~Betke, C.~Willett, G.~Chen, The correlation
  between internal and external markers for abdominal tumors: Implications for
  respiratory gating, International Journal of Radiation
  Oncology*Biology*Physics 61~(5) (2005) 1551 -- 1558
\newblock  (2005), \href
  {https://doi.org/https://doi.org/10.1016/j.ijrobp.2004.12.013}
  {\path{doi:https://doi.org/10.1016/j.ijrobp.2004.12.013}}.

\bibitem{RubioSBRT2013}
C.~Rubio, R.~Morera, O.~Hernando, T.~Leroy, S.~Lartigau, Extracranial
  stereotactic body radiotherapy. review of main sbrt features and indications
  in primary tumors, Reports of Practical Oncology \& Radiotherapy 18~(6)
  (2013) 387 -- 396, selected Papers Presented at the XVII SEOR Congress, Vigo,
  18—21 June 2013
\newblock  (2013), \href
  {https://doi.org/https://doi.org/10.1016/j.rpor.2013.09.009}
  {\path{doi:https://doi.org/10.1016/j.rpor.2013.09.009}}.

\bibitem{HoogemanCyberKnifeClinicalAccuracy2009}
M.~Hoogeman, J.-B. Prévost, J.~Nuyttens, J.~Pöll, P.~Levendag, B.~Heijmen,
  Clinical accuracy of the respiratory tumor tracking system of the cyberknife:
  Assessment by analysis of log files, International Journal of Radiation
  Oncology*Biology*Physics 74~(1) (2009) 297 -- 303
\newblock  (2009), \href
  {https://doi.org/https://doi.org/10.1016/j.ijrobp.2008.12.041}
  {\path{doi:https://doi.org/10.1016/j.ijrobp.2008.12.041}}.

\bibitem{SamavatiHybridDIR4DCT2015}
N.~Samavati, M.~Velec, K.~Brock, A hybrid biomechanical intensity based
  deformable image registration of lung 4dct, Physics in Medicine \& Biology
  60~(8) (2015) 3359
\newblock  (2015).

\bibitem{BrockDIR2010}
K.~Brock, Results of a multi-institution deformable registration accuracy study
  (midras), International Journal of Radiation Oncology*Biology*Physics 76~(2)
  (2010) 583 -- 596
\newblock  (2010), \href
  {https://doi.org/https://doi.org/10.1016/j.ijrobp.2009.06.031}
  {\path{doi:https://doi.org/10.1016/j.ijrobp.2009.06.031}}.

\bibitem{WilmsImproved4DCT2017}
M.~Wilms, R.~Werner, T.~Yamamoto, H.~Handels, J.~Ehrhardt, Subpopulation-based
  correspondence modelling for improved respiratory motion estimation in the
  presence of inter-fraction motion variations, Physics in Medicine {\&}
  Biology 62~(14) (2017) 5823--5839
\newblock  (jun 2017), \href {https://doi.org/10.1088/1361-6560/aa70cc}
  {\path{doi:10.1088/1361-6560/aa70cc}}.

\bibitem{FayadNo4DCTrequired2018}
H.~Fayad, M.~Gilles, T.~Pan, D.~Visvikis, A 4d global respiratory motion model
  of the thorax based on ct images: A proof of concept, Medical Physics 45~(7)
  (2018) 3043--3051
\newblock  (2018), \href {https://doi.org/10.1002/mp.12982}
  {\path{doi:10.1002/mp.12982}}.

\bibitem{McClellandReconstructionFromPartialData2017}
J.~R. McClelland, M.~Modat, S.~Arridge, H.~Grimes, D.~D'Souza, D.~Thomas, D.~O.
  Connell, D.~A. Low, E.~Kaza, D.~J. Collins, M.~O. Leach, D.~J. Hawkes, A
  generalized framework unifying image registration and respiratory motion
  models and incorporating image reconstruction, for partial image data or full
  images, Physics in Medicine and Biology 62~(11) (2017) 4273--4292
\newblock  (may 2017), \href {https://doi.org/10.1088/1361-6560/aa6070}
  {\path{doi:10.1088/1361-6560/aa6070}}.

\bibitem{SeregniArtificialNeuralNetworks2013}
M.~Seregni, A.~Pella, M.~Riboldi, R.~Orecchia, P.~Cerveri, G.~Baroni, Real-time
  tumor tracking with an artificial neural networks-based method: A feasibility
  study, Physica Medica 29~(1) (2013) 48 -- 59
\newblock  (2013), \href
  {https://doi.org/https://doi.org/10.1016/j.ejmp.2011.11.005}
  {\path{doi:https://doi.org/10.1016/j.ejmp.2011.11.005}}.

\bibitem{IsakssonAdaptiveNeuralNetwork2005}
M.~Isaksson, J.~Jalden, M.~Murphy, On using an adaptive neural network to
  predict lung tumor motion during respiration for radiotherapy applications,
  Medical Physics 32~(12) (2005) 3801--3809
\newblock  (2005), \href {https://doi.org/10.1118/1.2134958}
  {\path{doi:10.1118/1.2134958}}.

\bibitem{ZhangNearestNeighborMethod2018}
J.~Zhang, X.~Huang, Y.~Shen, Y.~Chen, J.~Cai, Y.~Ge, Nearest neighbor method to
  estimate internal target for real-time tumor tracking, Technology in Cancer
  Research \& Treatment 17 (2018) 1533033818786597, pMID: 30081745
\newblock  (2018), \href {https://doi.org/10.1177/1533033818786597}
  {\path{doi:10.1177/1533033818786597}}.

\end{thebibliography}

\end{document}